\documentclass[preprint, 11pt,longtitle]{elsarticle}
\usepackage{graphicx}
\usepackage{color}
\usepackage{amssymb}
\usepackage{amsfonts}
\usepackage{mathtools}
\usepackage{tikz} 
\usepackage{pgfplots}
\usepackage{geometry}
\usepackage{setspace}
\usepackage{caption}
\usepackage{subcaption}
\usepackage{lipsum}
\usepackage{wrapfig}
\usepackage{float}
\usepackage{todonotes}
\usepackage{bm,nicefrac}
\usepackage{listings}
\usepackage{algorithm2e}
\usepackage{extarrows}
\usepackage{enumitem}
\setlist[enumerate]{itemsep=0mm}

\newtheorem{lem}{Lemma}

\newtheorem{prop}{Proposition}

\newtheorem{Th}{Theorem}[section]

\newcommand{\ovl}{\overline}

\newcommand{\ignore}[1]{}

\newcommand{\PP}{\mathbb{P}}

\newcommand{\Q}{\mathcal{Q}}

\title{Combinatorial results for network-based models of metabolic origins}
\author{Oliver Weller-Davies$^a$, Mike Steel$^b$ and Jotun Hein$^c$}
\date{May 25, 2019}

\begin{document}

\begin{abstract}
A key step in the origin of life is the emergence of a primitive metabolism. This requires the formation of a subset of chemical reactions that is both self-sustaining and collectively autocatalytic. A generic theory to study such processes (called `RAF theory') has provided a precise and computationally effective way to address these questions, both on simulated data and in laboratory studies. One of the classic applications of this theory (arising from Stuart Kauffman's pioneering work in the 1980s) involves networks of polymers under cleavage and ligation reactions;  in the first part of this paper, we provide the first exact description of the number of such reactions under various model assumptions.  Conclusions from earlier studies relied on either approximations or asymptotic counting, and we show that the exact counts lead to similar (though not always identical) asymptotic results.  In the second part of the paper, we solve some questions posed in more recent papers concerning the computational complexity of some key questions in RAF theory. In particular, although there is a fast algorithm to determine whether or not a catalytic reaction network contains a subset that is both self-sustaining and autocatalytic (and, if so, find one), determining whether or not sets exist that satisfy certain additional constraints exist turns out to be NP-complete. \end{abstract}

\bigskip

\address{ \textsuperscript{a} Department of Statistics, Oxford University, Oxford UK \\ {\em Email:}  oliver.weller-davies@keble.ox.ac.uk (corresponding author).
}
\address{\textsuperscript{b}Biomathematics Research Centre, University of Canterbury, Christchurch, New Zealand.
{\em Email:}  mike.steel@canterbury.ac.nz
}
\address{\textsuperscript{(c)} Department of Statistics, Oxford University, Oxford UK
{\em Email:}  hein@stats.ox.ac.uk
}
\maketitle


\newpage

\section{Introduction}
\label{sec:intro}

The origin of life remains an unsolved challenge in science. Once considered a problem `beyond science', many researchers now believe its solution may not be far off \cite{mar, vas12}. This  prospect has partly been fuelled by recent efforts to integrate formal models and mathematical techniques into the field.
Initially motivated by the emergent qualitative properties of discrete random networks  (such as early works of Paul Erd{\"o}s and  Alfred R\'enyi  \cite{erd60}),
 one approach to formally capture `life-like' emergence in a chemical system involves the notion of a collectively autocatalytic sets, a concept pioneered for polymer systems by Stuart Kauffman \cite{kau86}. This approach was  later developed more formally as the notion of  \textit{Reflexively Auto-catalytic F-generated sets} (RAFs). Such sets couple together two basic requirements for any living system: reactions are catalysed by molecules types (e.g. enzymes, cofactors etc) generated from within the system, and second, every reaction within the system requires just molecule types that can be constructed from an ambient food source by using just reactions within the system (i.e. it is self-sustaining from the chemistry of the environment). 
 
  In a series of papers beginning in 2000 to the present  \cite{hei10, hor04, ste00}, RAFs have been investigated in both simulated and laboratory-based systems of early metabolism (\cite{ash04, vai12} as discussed in \cite{ste18}) as well as in
the analysis of metabolism in the bacterium {\em Escherichia coli} \cite{sou15} and a recent study into ancient metabolism revealed by analysing large biochemical databases \cite{xav19}. RAF theory has also been applied to a variety of different settings and scenarios related to the origin and organisation of life and metabolism, as well as some applications in other fields such as ecology \citep{caz18}, economics  \citep{hor17} and cultural evolution \citep{gab17}. 

In this paper, we derive a number of new results in RAF theory, answering some outstanding complexity questions that have been posed in earlier papers. We also provide a more exact treatment of the enumeration of reactions in polymer models than has been given in any previous work.  We begin in Section~\ref{sec:def},  by providing basic definitions and summarizing some results and questions from RAF theory.  In Section~\ref{sec:poly} we review existing polymer models (based on `oriented' polymers), and then investigate the consequences of variations on this model that have been overlooked in earlier work, as well as derive precise enumerative and asymptotic  formulae for the number of reactions in these cases. We then explore another polymer model (`non-oriented') that does not appear to have been considered before in the RAF setting. We again  derive exact and asymptotic formulae, and the consequences of these are also briefly discussed.  

In Section~\ref{sec:closed},  we investigate the computational complexity of  two problems (posed in \cite{hor18} and \cite{ste18}) concerning RAFs that are `closed' (a biochemically relevant condition we describe there). Namely, (i) does a RAF set contain a closed subRAF set?   (ii) Is there a closed RAF set that is `uninhibited' (i.e., no molecule inhibits any reaction) in the simple case when (just) a single molecule inhibits (just) one reaction?  We show that both questions are NP-complete (even though Question (i) is easily solved without the `closure' restriction). 

In Section~\ref{sec:compl},  we investigate the simpler `elementary' CRS framework (where the reactants of each reaction are present in the food source) and solve two  complexity problems (posed in \cite{sou15}, \cite{ste18}): (i) What is the size of the largest irreducible RAF?   (ii) Does an uninhibited RAF exist?  We show that these questions are NP-hard in this elementary CRS setting.
 
 We end with some brief concluding comments in Section~\ref{sec:conc}.

\section{Definitions and mathematical preliminaries}
\label{sec:def}

RAFs are defined within the context of a \textit{catalytic reaction system}. Formally, a catalytic reaction system (CRS) is a quadruple $\Q = (X, R, C, F)$ where:
\begin{itemize}
	\item $X$ denotes a set of \textit{molecule types}.
	\item $R$ denotes a set of reactions between sets of molecule types. Specifically, each $r = (A, B) \in R$ is a pair of \textit{sets} of molecule types $A, B \subseteq X$. The molecule types in $A$ are referred to as the \textit{reactants} of the reaction; those in $B$ as the \textit{products} of the reaction.
	\item $C \subseteq X \times R$ denotes a \textit{catalysation} assignment; where if $(x, r) \in C$, we say that the molecule $x$ \textit{catalyses} the reaction $r$.
	\item $F \subseteq X$ denotes an ambient \textit{food set} of molecule types, which are assumed to be freely available in the environment.
\end{itemize}

A simple example of a CRS is the binary polymer model introduced by   \cite{kau86}.  Suppose, under appropriate chemical conditions, that we have a set of polymeric molecules.  Here, `polymeric' refers to  macromolecules formed from the repeated concatenation of sub-molecules (monomers), such as how the polymer RNA is formed by repeated joins of the bases (monomers) A,U,C,G, or how proteins are formed from amino acids.
Polymers can be  represented as  strings which leads naturally to two types of reaction. The first type, ligation reactions, denote the concatenation of two polymers (strings) to form a longer polymer. For example, in the binary (a,b) polymer setting, we have:  $$baba+ bbb \xrightarrow{} bababbb \text{  (ligation reaction)}.$$ The second type of reaction (cleavage) is the reverse of ligation; it denotes the splitting up of a polymer into appropriate constituent sub-strings; for example: $$bababbb \xrightarrow{} baba + bbb \text{  (cleavage reaction)}.$$
We fix a constant $n$ to denote the maximum length of polymers in the system. Within this setting, we also assume that all polymers of length $\leq t$ are freely available and constantly replenished (a food set). Catalysis is typically modelled as a stochastic process -- where each polymer  catalyses a given ligation/cleavage reaction with a certain probability. 

\begin{figure}[ht]
\begin{center}
\includegraphics[width=0.5\linewidth]{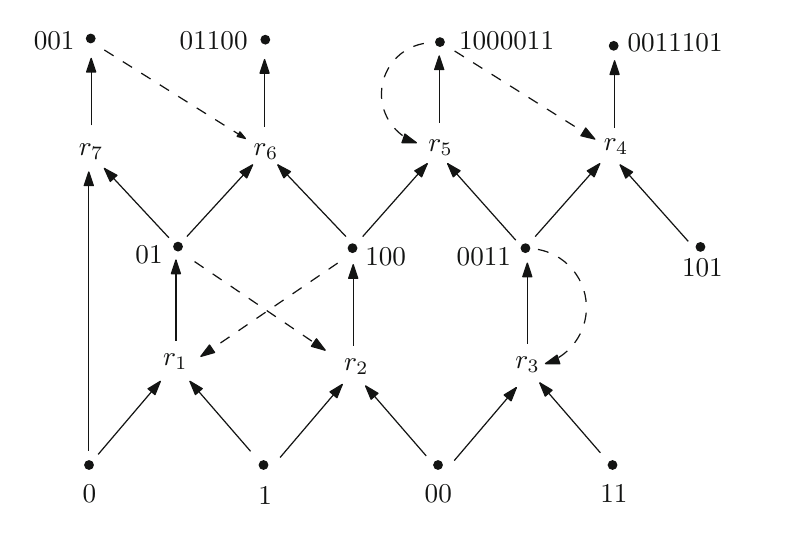} 
\caption{A simple CRS based on polymers over a two-letter alphabet (0,1), with a food set
$F = \{0, 1, 00, 11\}$ and seven reactions ($r_1$ -- $r_7$) (from  \cite{smi14}).  Dashed arrows indicate catalysis; solid arrows show reactants entering a reaction and products leaving. In this CRS, there are exactly four RAFs (defined below), namely $\{r_1, r_2\}, \{r_3\}, \{r_1, r_2, r_3\}$ and $\{r_1,r_2,r_3, r_5\}$.}
\label{fig1}
\end{center}
\vspace{-10pt}
\end{figure}

Any  CRS (not only polymer models) can be represented as a directed graph with two types of vertices (molecule types and reactions) and  two types of directed edges, one showing the flow of molecule types into and out of reactions, and the other type being catalysation of a reaction by a molecule type. When drawing these graphs, we typically follow the convention that molecule types are represented by circles, reactions by squares, reactant and product directed edges by solid-line arrows and catalysation directed edges by dashed-lines arrows.  See Fig.~\ref{fig1}.

Given a CRS $Q=(X, R, C, F)$, a subset of reactions $R' \subseteq R$ is a {\em RAF} set for $\Q$
if $R'$ is non-empty and both of the following conditions hold:
\begin{itemize}
    \item[(RA)] \textit{Reflexively Auto-catalytic:} {Every reaction $r \in R'$ is catalysed by a molecule type $x$ that is either in the food set  $F$ or is the product of another reaction $r' \in R'$.}
    \item[(F)] \textit{F-generated}: {The reactions in $R'$ can be written in a linear order $r_0, r_1,\ldots,r_n$ such that for every reaction $r_i = (A_i, B_i) \in R'$, each reactant $x \in A_i$ is either in the food set or is the product of another reaction occurring earlier in the ordering; that is, $\forall x \in A_i$, $x \in F$ or $x \in B_j$ for some $j < i$.}
\end{itemize}

\begin{figure}[ht]
\begin{center}
\includegraphics[width=0.7\linewidth]{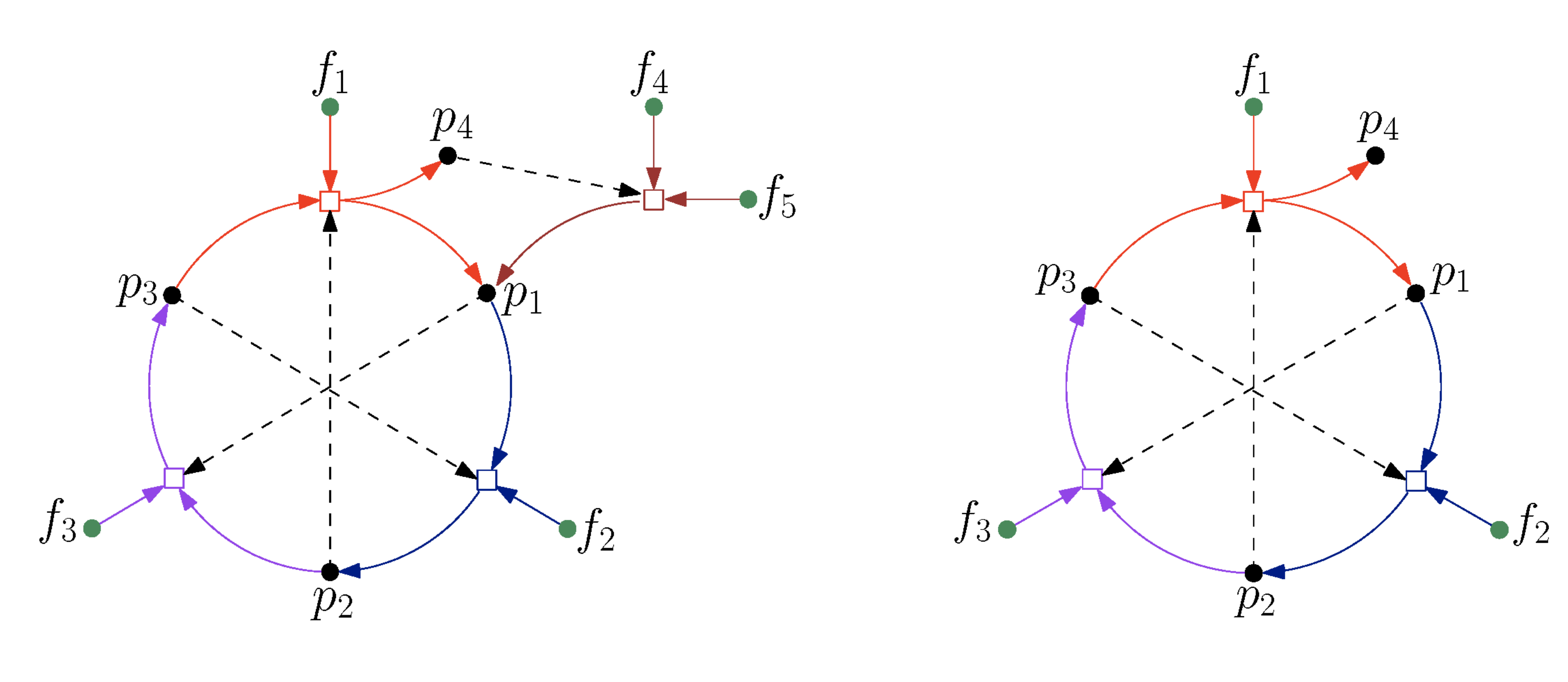} 
  \caption{The set in (a) is an example of a  RAF, while (b) is not a RAF (since it fails to be $F$-generated). Here, food is indicated by $f_*$ (green vertices) and catalysis by dashed arrows.}
  \label{fig2}
\end{center}
\end{figure}

Roughly speaking, a RAF set is a subset of reactions that is both collectively autocatalytic (i.e. each reaction is catalysed by some molecule type involved in the reaction subset) and self-sustaining (i.e. each molecule type required for the reactions to occur can be built up starting from just the food set; using reactions only within the reaction subset). 
The RAF concept is illustrated in Fig.~\ref{fig2}(a), along with the weaker notion of a `pseudo-RAF' (Part (b) of Fig.~\ref{fig2}) that satisfies the (RA) condition but fails to be $F$-generated.

RAFs underly the metabolism of both existing cellular life \cite{sou15}  and also arise in  laboratory models of early life  \cite{ash04, ste18, vai12}.
The concept of a RAF couples two features that seem to be essential in the earliest metabolism at the origin of life, as well as in extant cellular life.
Firstly, the reactions need to be catalysed by molecules present in the system (in modern metabolism, these catalysts are highly efficient, and are based on enzymes and cofactors, whereas in early metabolism, it is likely that much simpler catalysts based on metals such as iron would more likely be involved). Biochemical catalysts not only speed up reactions by many orders of magnitude but they also allow reactions to be synchronised \cite{wol01}.  Secondly, the system must be `self-sustaining' from an available (external) food source;   in other words, the reactants of each reaction in the system must either be in the food set or able to be built up from the food set by a sequence of reactions within the system. These two features are combined into the two conditions (RA) and (F). 

It is easily seen that the union of two or more RAFs is a RAF, and thus if a CRS has a RAF, it has a unique maximal one, called the maxRAF.  Although it is not perhaps obvious from the above definition, it turns out that there is a (polynomial-time) algorithm for computing the {\rm maxRAF} within a given CRS, if one exists at all \cite{hor04}. We describe this maxRAF algorithm shortly.
Any given RAF $R'$ may contain another RAF $R''$ as a strict subset, in which case we say that $R''$ is a {\em subRAF} of $R'$.


\subsection{The closure of a set of molecule types}

Fix a CRS $\Q = (X, R, C, F)$. Given a set of reactions $R' \subseteq R$ and a set of molecule types $X' \subseteq X$, the molecule closure ${\rm cl}_{R'}(X')$ is the unique minimal subset $W \subseteq X$ of molecule types satisfying the following two conditions: (i) $X' \subseteq W$;  (ii) for every reaction $(A, B) \in R'$: $A \subseteq W \implies B \subseteq W$ (as defined in \cite{hor04}).

In words, ${\rm cl}_{R'}(X')$ denotes the set of molecule types arrived at if we were to continually apply reactions from $R'$, wherever we could, ignoring catalysis constraints and starting only with molecule types from $X'$. Mostly we will be considering the case where $X'$ is the food set $F$.  To aid the mathematical analysis, we will use an equivalent but alternative definition of a RAF set (as defined and justified in \cite{ste13}) that incorporates the notion of molecule closure. 

\begin{lem}
Given a CRS  $\Q= (X, R, C, F)$, a subset of reactions $R' \subseteq R$ is a RAF for $\Q$  if and only if  it is non-empty and both of the following conditions hold:
\begin{itemize}
	\item[(i)] For every reaction $r \in R'$, there is at least one molecule type $x \in {\rm cl}_{R'}(F)$ with $(x, r) \in C$;
	\item[(ii)] For every reaction $r = (A, B) \in R'$, $A\subseteq {\rm cl}_{R'}(F)$.
\end{itemize}
\end{lem}

Condition (ii) is equivalent to the F-generated condition, but Condition (i) is stronger than the RA condition described earlier; however, the combination of Conditions (i) and (ii) are equivalent to the combination of the earlier conditions of RA and F-generated. 


Given a CRS, $\Q=(X, R, C, F)$,  the  {\rm maxRAF} algorithm computes a nested decreasing sequence of subsets of $R$, starting from $R$:
 $$R=R_0 \supset R_1 \supset \cdots \supset R_k = R_{k+1}$$
 where $R_{j+1}$ is the subset of reactions in $R_j$ that have all their reactants and at least one catalyst in ${\rm cl}_{R_j}(F)$.
At the first value of $k$ for which  $R_k=R_{k+1}$,  this set is either empty (in which case, $\Q$ has no RAF)  or it is the unique maximal RAF set (${\rm maxRAF}$).  For a proof of these assertions, see \cite{hor04}. A rudimentary runtime analysis of the gives us $O(|R|^3|X|)$ time and $O(|X| + |R|)$ space \cite{hor04}, although optimised implementations exist which tend to run sub-quadratically (in time) with the number of reactions \cite{hor15}.
\begin{figure}[ht]
\begin{center}
\includegraphics[width=1.0\linewidth]{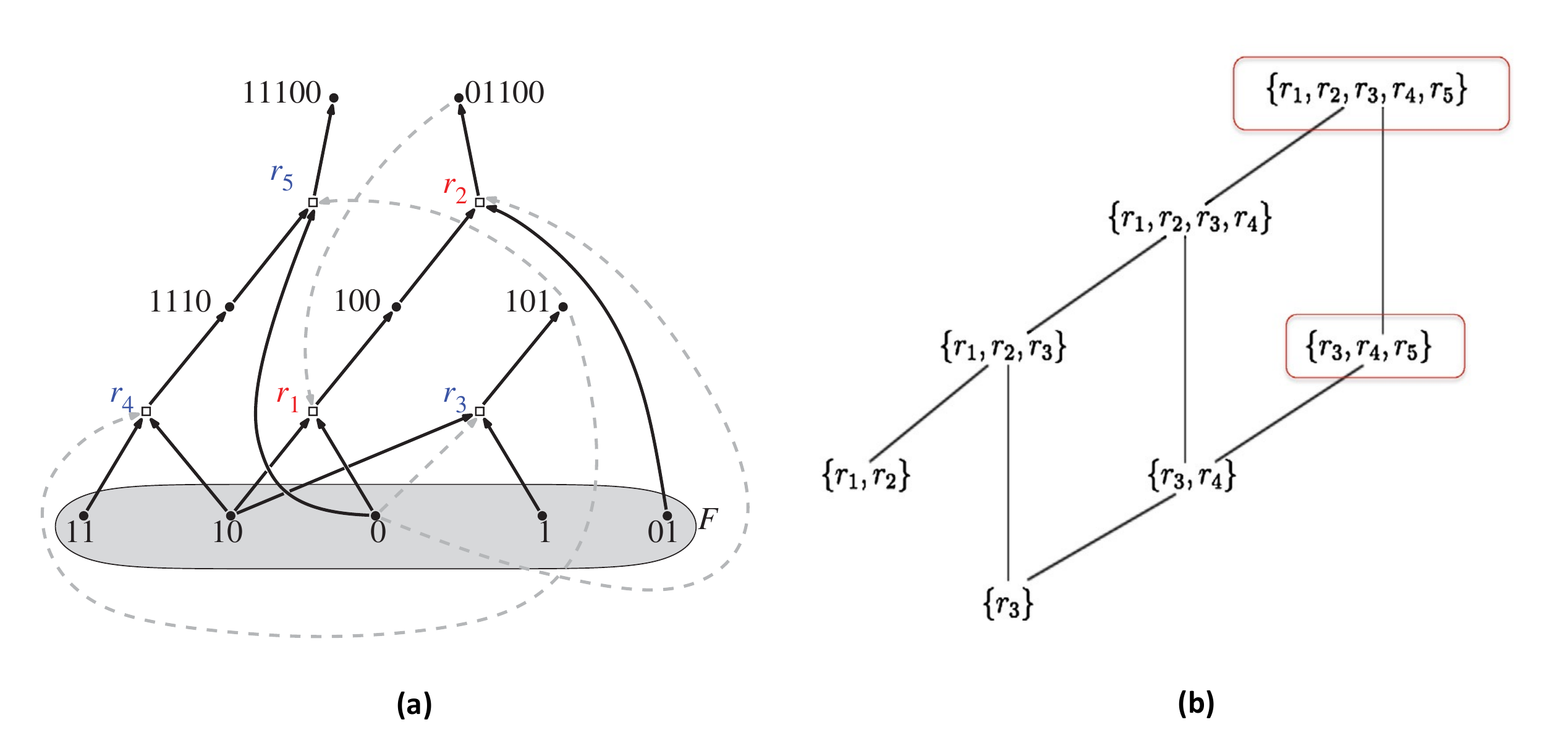}
  \caption{{\em Left:}  The CRS above is a portion of the binary polymer model instance with food molecule types highlighted in the lower shaded section. The maxRAF of the CRS contains all reactions $\{r_1, r_2, r_3, r_4, r_5\}$. {\em Right:} The partially ordered set (poset) of RAFs for the CRS. The two circled RAFs are closed (a notion defined and studied in Section 4).}
  \label{fig3}
\end{center}
\end{figure}


\section{Polymer models: Combinatorial enumeration and its role in RAF emergence}
\label{sec:poly}

We now describe polymer models more formally, following \cite{mos05, ste00}, based on earlier work by  \cite{kau86}.
In particular, we provide a mathematical analysis of some hitherto overlooked subtleties 
and variations concerning the classification and counting of molecule types and reactions within polymer models. The reason why such enumeration is important is that the key mathematical results concerning the emergence of RAFs in polymer models when the catalysis rates pass a certain threshold depend ultimately on the number of reactions and molecule types, and the ratio of these two quantities.  We describe this relevance in Section~\ref{relesec}.

We first describe the set of molecule types in polymer systems. Consider  an arbitrary finite alphabet $\Sigma$ of size $k \geq 2$ and let $\Sigma^+$ denote the set of all finite-length (ordered) words formed from this alphabet. For $w \in \Sigma^+$, we will let $|w|$ denote the length (number of characters) in $w$. We refer to $w$ as an {\em oriented polymer} over the alphabet $\Sigma$.

In the original polymer model from \cite{kau86}, developed further in \cite{mos05, ste00},  the set $X$ of molecule types consists of  oriented polymers over the alphabet $\Sigma$.
More precisely, for $n \geq 1$, let $X_n$ be the set of  words  $w \in \Sigma^+$ of length at most $n$ (i.e. $1 \leq |w|\leq n$).  In this case,
the number of different molecule types $|X_n| = |\{ w \in \Sigma^+ : |w| \leq n \}|$ is just the sum $\sum_{i = 1}^{n} k^i$ (all non-empty words of length $\leq n$) and therefore: 
\begin{equation}
|X_n| = \frac{k^{n+1} - k}{k-1} \sim \frac{k^{n+1}}{k - 1}
\end{equation}
where here and below $\sim$ denotes asymptotic equivalence as $n$ grows (we regard $k$ as fixed throughout).
In other words,  $f(n) \sim g(n)$ if and only if  $\lim_{n \rightarrow \infty}{f(n) / g(n) = 1}$ (for details, see \cite{mos05}).

\subsection{Reaction sets and their enumeration}  

In the polymer model, reactions consist of two complementary types. The first is a  {\em ligation}  reaction where two polymers are concatenated, formally:
$$w + w' \rightarrow ww'.$$
The second is a {\em cleavage} reaction where a polymer is split in two:
$$ww' \rightarrow w+w'.$$

Note that the cleavage and ligation reactions are reversals of each other and thus are in one-to-one correspondence.  We will therefore mostly concentrate on enumerating ligation reactions.

We now highlight a subtle but important distinction in the following example.
Consider the two cleavage reactions:
\begin{equation}
\label{ww1}
ab + abab \rightarrow  ababab
\end{equation}
and 
\begin{equation}
\label{ww2}
abab +ab \rightarrow ababab.
\end{equation}
These two reactions have the same reactants and the same product, and the only distinction is the order in which the reactants appear on the left. 
Thus it is tempting to regard (\ref{ww1}) and (\ref{ww2}) as the same reaction; we call this {\em Convention A}, and it was the one that was assumed in \cite{mos05, ste00}.  However, given that the polymers are oriented,
one might regard the first reaction above as attaching $ab$ to the left-hand end of the (oriented) polymer $abab$ whereas the second reaction is attaching $ab$ to the right-hand end of the
polymer $abab$; in this way, the reactions can regarded as different. We call this {\em Convention B}, and it has also been tacitly assumed in other papers on the studies involving the model, particularly with simulations.    Note that this distinction between these two conventions vanishes in the non-oriented setting.

Let  $R^A_n$ and $R^B_n$ denote the set of ligation reactions involving oriented polymers of size less or equal to  $n$ under Conventions A and B, respectively. 
The one-to-one correspondence between  cleavage reactions  and  ligation reaction holds both for Convention A and for Convention B.
Calculating the size of $R^B_n$ is easy and was carried out in the earlier papers cited above. 
For $i\geq 2$, if we let $S(i)$ denote  the number of ordered triples of oriented polymers $(u,v,w)$ where $uv=w$ and $|w|=i$, 
then:
\begin{equation}
\label{eqs}
S(i) = (i-1)k^i,
\end{equation}
and we have:
\begin{equation} 
\label{twoass}
|R^B_n| = \sum_{i=1}^n S(i) = \frac{nk^{n+1}}{k-1}  - \frac{k^2(k^n-1)}{(k-1)^2}  \sim  \frac{nk^{n+1}}{k-1}.
\end{equation}

For Convention A, the expression for $|R^A_n|$ is more delicate. To describe this, fix an alphabet $\Sigma$ of size $k$, and let $U(i)$ denote the number of unique unordered sets $\{u,v,z\}$ for $u,v,z \in {\Sigma}^+$ with $uv = z$ and $|z|= i$.
 We first state a lemma, which is a consequence of a more general result from Lyndon and Sch{\"{u}}tzenberger \cite{lyn62}.
\begin{lem}
\label{abba}
If $uv = vu$ for non-empty words $u$ and $v$, then $u$ and $v$ are powers of a common word; that is, there exists a word $w$ such that $w^i = u$ and $w^j = v$ for some $i,j \geq 1$. Therefore, $uv = w^{i+j}$.  Here, $w^i$ denotes $i$ repeated copies of $w$ joined together.
\end{lem}
\hfill$\Box$

This lemma allows us to characterise strings of the form $uv = vu
$ with $u, v \in \Sigma^+$. In particular, if $z$ is a word of the form $z = uv=vu$ for $u,v \in \Sigma^+$,  then $z$ is necessarily a word formed from a repeated sub-word. Words that can be decomposed in such a way are often referred to as \textit{periodic}, where their \textit{period} is taken to be the  smallest  sub-word that repeatedly joins to form the word; for instance, the word $abcabcabcabc$ has period $abc$. The period of a word, if it exists, must be unique. Conversely, words that cannot be formed from repeated substrings in such a way are often referred to as \textit{aperiodic} or \textit{primitive} words (e.g. $abcdef$ is aperiodic).

This leads to the following expression for $|R^A_n|$ (details of the proof of Part (i) are in the Appendix), where 
 $\mu$ denotes the (classical) M\"obius function for the partial ordered set of positive integers under division  \cite{sta11}, defined by:

\begin{equation}
\label{mobiii}
    \mu(x) = 
    \begin{cases*}
      1, & if $x = 1;$ \\
      0, & is $x$ is a square number; \\
      (-1)^r, & if $x$ is square-free with $r$ distinct prime factors. \\
    \end{cases*}
  \end{equation}

\begin{Th}
\label{main1}
Fix an alphabet $\Sigma$ of size $k$.  
Let $|R^A_n|$ denote the set of cleavage and ligation reactions in the oriented polymer model  under Convention A, with an alphabet  of size $k$ and polymer length of at most $n$. 
We then have the following:
\begin{itemize}
\item[(i)] 
$|R^A_n|=\sum_{i=2}^n U(i)$, where:
\begin{equation} \label{rMob}
U(i)=
    (i-1)k^i - \sum_{d|i, d<i}
    {
      {\left \lfloor \frac{i/d - 1}{2}\right \rfloor}
      \sum_{d'|d}{
          \mu\left(\frac{d}{d'}\right)k^{d'}
      }
    }
\end{equation}
  
  \item[(ii)]
$S(i) - U(i) = O(k^{i/3})$, where $S(i)$ is given in (\ref{eqs}),
and so $|R^A_n| \sim |R^B_n|$ as $n$ grows.
\end{itemize}
\end{Th}

\begin{figure}[ht]
\begin{center}
\includegraphics[width=0.55\linewidth]{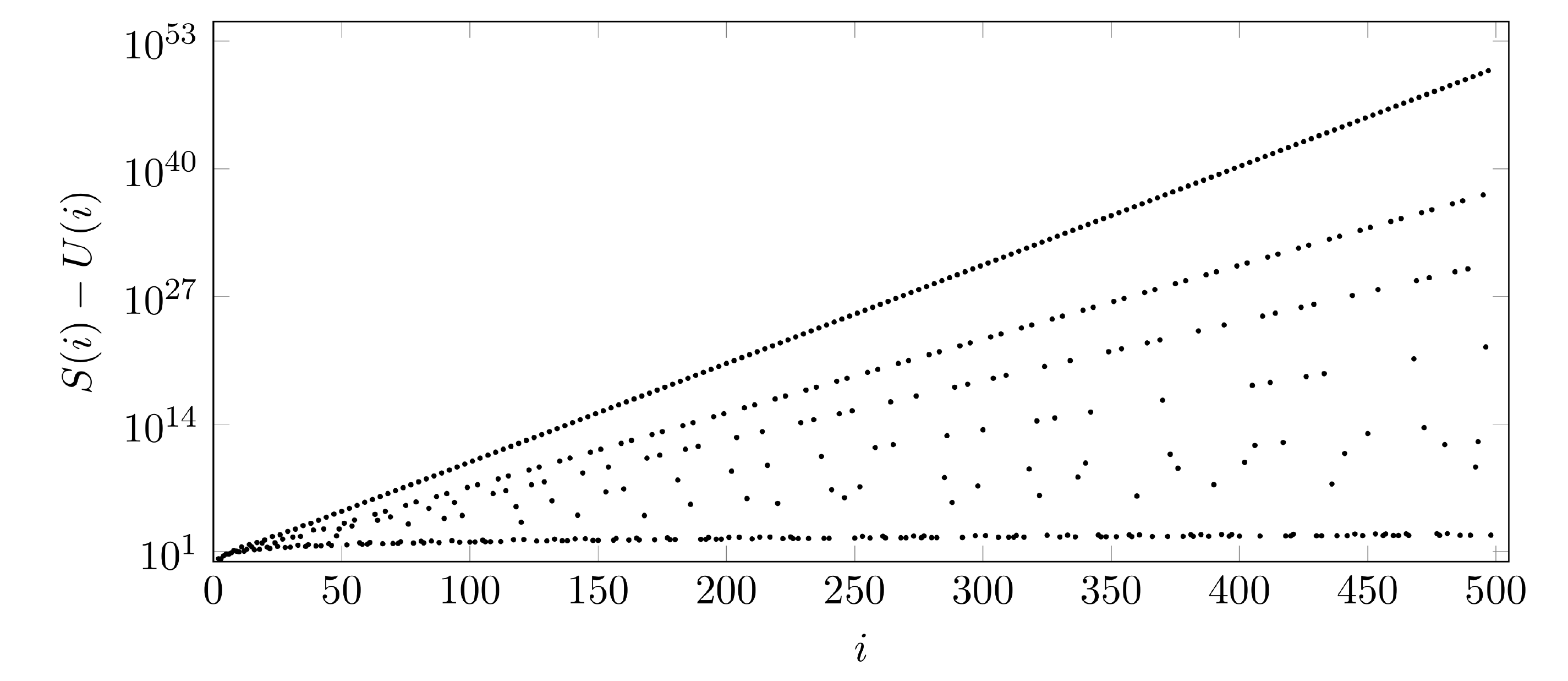} 
  \caption{The difference between $S(i)$ and $U(i)$ as $i$ increases from 1 to 500}
    \label{fig4}
\end{center}
\vspace{-12pt}
\end{figure}

To illustrate Theorem~\ref{main1}, consider the binary polymer model (i.e. $k=2$). By Part (i) of Theorem~\ref{main1} we have $U(2)=S(2)=4$ and $U(3) = 2.2^3 - 2 = 14$, while $S(3)= 16$. Thus  $|R_3| = U(2)+U(3) = 18$.  In this case, for the two reaction pairs:
$$aa + a \rightarrow aaa, \mbox{  } a + aa\rightarrow aaa$$ 
$$bb + b \rightarrow bbb, \mbox{ } b+ bb \rightarrow bbb,$$ 
the reactions within each pair are counted separately by $S(3)$ but only once by $U(3)$. The  difference  $S(i) - U(i)$ quantifies the difference between Conventions A and B when the product of a cleavage reaction has size $i$. A graph of this difference is shown in Fig.~\ref{fig4}. 

{\em Outline Proof of Theorem~\ref{main1} :} The proof of Part (i) is given in the Appendix. 
For Part (ii), Lemma~\ref{abba} gives: $S(i)-U(i)$ is the number of pairs of words of the form  $w^l, w^m$ where  $1 \leq l, m \leq i/|w|$ satisfies $(l+m)|w|=i$ and $l < m$.
Let $r:=l+m$ (so $|w| = i/r$). Since $l, m \geq 1$ and $l\neq m$, we have $r \geq 3$.  Now if $|w|=i/r$ there are $k^{i/r}$ choices for $|w|$ and so
 $S(i)-U(i)$ is bounded above, as claimed,  by: $$\sum_{r=3}^i (r-1) k^{i/r} = \Theta(k^{i/3}).$$
The second claim in  Part (ii) now follows by applying Part (i) and Eqn.~(\ref{twoass}) to give: $$|R^A_n| = \sum_{i=2}^n U(i) \sim \sum_{i=2}^n S(i).$$ The asymptotic equivalence here holds because the sum of terms of order $k^{i/3}$ from $i=1$ to $i=n$ is of order $k^{(n+1)/3}$ and thus is asymptotically negligible compared with $|R^A_n|$.
\hfill$\Box$

\bigskip

\subsection{Non-oriented polymers}
The choice to consider polymers as oriented is not only computationally convenient; it also has some connection with bio-molecular sequence data, where (for example) DNA polymerase has an orientation (from the 5$'$ to 3$'$ end of the sequence). However, it may also be useful, particularly in the study of early biochemistry, to regard polymers as non-oriented; in other words, a polymer such as $accb$ should be considered as identical to $bcca$, since one can rotate the first molecule in space to obtain the second. Thus, each oriented polymer $w$ can be viewed as equivalent to its `reverse' oriented polymer $w^{-}$ obtained by reversing the order of the letters in the polymer.
Note that a polymer is equal to its associated reverse polymer  precisely if it is a palindromic polymer (e.g. $acca$).  In this way, the set $\Sigma^+$ is 
partitioned into pairs and singleton classes under the equivalence relation $w \sim w'$ if $w' = w$ or $w'=w^{-}$. We refer to these equivalence classes as {\em non-oriented polymers} and we will write $\overline{w}$ for the equivalence class of $w$. 
Let $\overline{X}_n = \{\overline{w} : w \in \Sigma \text, |w| \leq n\}$ denote the set of non-oriented polymers (equivalence classes) for words $w \in \Sigma$ with $1\leq |w|\leq n$. 

\bigskip

\noindent\fbox{%
    \parbox{\textwidth}{%
\noindent {\em Example:} 

For $\Sigma= \{a,b\}$, we have:  $|\overline{X}_2| = 5$, since $$\overline{X}_2 = \{\{a\}, \{b\}, \{aa\}, \{bb\}, \{ab, ba\}\}.$$ 
Similarly, $|\overline{X}_3| = 11$, since: $$\overline{X}_3 = \{\{a\}, \{b\}, \{aa\}, \{bb\}, \{ab, ba\}, \{aaa\}, \{bbb\}, \{aba\}, \{bab\}, \{abb, bba\}, \{aab, baa\}\}.$$
By contrast, $|X_2|=6$ and $|X_3| = 14$.
}}

\begin{lem}
\label{lemolem}
$$|\overline{X}_n| = \begin{cases}
\frac{k}{2(k-1)} \left( k^n + 2k^{n/2} -3\right), & \mbox{ if $n$ is even}; \\
\frac{k}{2(k-1)}\left (k^n + k^{(n+1)/2} -1\right), & \mbox{ if $n$ is odd}.
\end{cases}
$$
In particular, $|\overline{X}_n| \sim \frac{1}{2} |X_n|$ as $n$ grows.
\end{lem}
{\em Proof:} Let $p_i$ (respectively, $n_i$) denote the number of palindromic (respectively non-palindromic) oriented polymers of length $i$.
We have:
$$|\overline{X}_n|  = \sum_{i=1}^n \frac{1}{2} n_i  + \sum_{i=1}^n p_i = \frac{1}{2} \left(\sum_{i=1} k^i + \sum_{i=1}^n p_i\right),$$
where the second equality follows from the identity:
 $n_i = k^i - p_i$. 
Now,  $p_i$ is equal to  $k^{i/2}$ when  $i$ is even and is equal to $k^{(i-1)/2} \times k = k^{(i+1)/2}$ when $i$ is odd.
Applying the geometric sequence identity
$\sum_{j=1}^m x^j = x(x^{m} -1)/(x-1)$ establishes the first claim. The second claim follows by combining this result with Eqn.~(\ref{twoass}). 
\hfill$\Box$

\bigskip

Next, we consider the enumeration of  involving non-oriented polymers. 
Consider the following  ligation reaction: 
\begin{equation}
\label{rr}
r: \overline{u} + \overline{v} \rightarrow \overline{z},
\end{equation} 
where $u, v$ and $z$ are oriented polymers.
Note that, for any two oriented polymers $x,y$ we always have $(xy)^- = y^- x^-$ and  the reaction $r$ in (\ref{rr})) holds whenever  $z$ is one of the following:
$$z = uv \mbox{ } (= (v^-u^-)^-),$$ 
$$z = uv^-\mbox{ }  (= (vu^-)^-),$$ 
$$z = u^-v \mbox{ }  (=(v^-u)^-),$$ 
$$z = u^-v^- \mbox{ }  (=vu)^-).$$
Thus $\overline{u}$ and $\overline{v}$ could ligate to form up to four different non-oriented polymers, depending on how many of these four possible $z$ are equivalent to each other (allowing reversals). 
There are four cases to consider:
\begin{itemize}
\item[(i)]
$u$ and $v$ are non-palindromic polymers (and $u \neq v, v^{-}$). In this case, there are four distinct choices of $z$ (no two of which are a reversal of each other),  so we obtain four distinct reactions of the type in (\ref{rr}).
\item[(ii)]If  $u,v$ are both non-palindromic and $u=v$ or $u=v^-$  we obtain three distinct choices of $z$ and so two additional distinct reactions of the type in (\ref{rr}). 
\item[(iii)] If one of $u$ or $v$ is a palindromic polymer and the other is not, 
we obtain two distinct choices of $z$ and so two distinct reactions of the type in  (\ref{rr}).
\item[(iv)] If both $u$ and $v$ are palindromic polymers, then all choices of $z$ are equivalent and so there is just one reaction of the type in (\ref{rr}).
\end{itemize}

\noindent\fbox{%
    \parbox{\textwidth}{%
\noindent {\em Example:}   

As an example of Case (i), consider the reaction of type (i):
$\overline{ab} + \overline{baa} \rightarrow \overline{abbaa}.$
The reactants on the left, give rise to three additional distinct reactions, namely:
$$\overline{ab} + \overline{baa} \rightarrow \overline{abaab}, \mbox{ } 
\overline{ab} + \overline{baa} \rightarrow \overline{babaa}, \mbox{ and } 
\overline{ab} + \overline{baa} \rightarrow \overline{baaab}.$$

Next, consider  the reaction of Type (ii):
$\overline{ab} + \overline{ba} \rightarrow \overline{abba}.$
The reactants on the left also give rise to: 
$$\overline{ab} + \overline{ba} \rightarrow \overline{baba} \mbox{ and }  \overline{ab} + \overline{ba} \rightarrow \overline{baab}.$$

Next, consider the reaction of Type (iii):
$\overline{aa} + \overline{ab} \rightarrow \overline{aaab}.$
The reactants on the left give rise to one additional reaction, namely 
$\overline{aa} + \overline{ab} \rightarrow \overline{aaba}.$

Finally,  the reaction of Type (iv)
$\overline{aa} + \overline{bab} \rightarrow \overline{aabab}$
gives rise to no further reactions. 

}
}
\bigskip

By applying these cases, once can,  in principle, count the number $N_{r,s}$ of cleavage reactions in which two non-oriented polymers of size $r$ and $s$ are combined to give a non-oriented polymer of length $i=r+s$. 
First, suppose that  $r<s$, so that Case (ii) cannot arise. We then have:
$$N_{r,s}= (2n_r + p_r)\cdot (2n_s+ p_s),$$
where, as before, $n_m$ (respectively, $p_m)$)  is the number of non-palindromic (respectively, palindromic) oriented polymers of length $m$.

The case where $r=s$ (in which case, $i= 2r$ is even) requires a separate description, since Case (ii) can now also arise. 
We have the following identity for the case $r=s$:

$$N_{r,r}=4n_r(n_r-1) + 3n_r+ 2n_rp_r + p_r(p_r+1)/2.$$

\noindent\fbox{%
    \parbox{\textwidth}{%
\noindent {\em Example:}   

Consider the case  $r=s=2$.  We have $n_2=1, p_2=2$ and so $N_{2,2}=10$.
These correspond to the following reactions, classified according to the cases described above. For Case (i) we have
 $4n_2(n_2-1)=0$, (i.e. no reactions are possible in this case).

For Case (ii),  $3n_r=3$:  $\overline{ab}+\overline{ab} \rightarrow \overline{abba},\mbox{ } \overline{ab}+\overline{ab} \rightarrow \overline{abab},\mbox{  } \overline{ab}+\overline{ab} \rightarrow \overline{baab}.$

For Case (iii), $2n_2p_2 = 4$:
$$\overline{aa}+\overline{ab} \rightarrow \overline{aaab},  \mbox{ } \overline{aa}+\overline{ab} \rightarrow\overline{aaba},$$
$$\overline{bb}+\overline{ab} \rightarrow \overline{abbb},  \mbox{ } \overline{bb}+\overline{ab} \rightarrow \overline{babb}$$

For Case (iv),  $p_2(p_2+1)/2 = 3$:
$\overline{aa}+\overline{aa} \rightarrow \overline{aaaa}, \mbox{  }  \overline{bb}+\overline{bb} \rightarrow \overline{bbbb}, \mbox{  } \overline{aa}+\overline{bb} \rightarrow \overline{aabb}.$

    }%
}

\bigskip

The total number of ligation reactions  in the non-oriented setting in which the product polymer has size exactly $i$  (respectively, has size at most $n$) 
is then $\sum_{r= 1}^{\lfloor i/2\rfloor}N_{r,i-r}$ (respectively, $\sum_{i= 2}^n \sum_{r= 1}^{\lfloor i/2\rfloor}N_{r,i-r}$). 

In principle, this provides an explicit (albeit complicated) expression for the total number of reactions; 
here, we just report the asymptotic behaviour as $n$ grows. Since the number of cleavage reactions is equal to the number of ligation reactions in the non-oriented model, it suffices to consider just cleavage reactions. We  have the following result: if we compare the non-oriented case to the oriented, the number of reactions  is (asymptotically) double and the ratio of reactions to molecule types is increased by a factor of 4.

\begin{prop}
\label{prol}
Let $\overline{R}_n$ denote the set of ligation reactions involving non-oriented polymers up to size $n$.
As $n$ grows and with $*=A, B$, we have:
$|\overline{R}_n| \sim 2 |R^*_n|$
and $|\overline{R}_n| /|\overline{X}_n| \sim 4 |R^*_n|/|X_n|$. 
\end{prop}
{\em Proof:}  We have $N_{r, i-r} = 4k^{i}-O(k^{i/2})$, from which it follows that:
$$\sum_{i= 2}^n \sum_{r= 1}^{\lfloor i/2\rfloor}N_{r,i-r} \sim \frac{2nk^{n+1}}{k-1}.
$$
The second result then follows from Lemma~\ref{lemolem}.

\subsection{Consequences for the emergence of RAFs in polymer systems}
\label{relesec}

We now turn to the relevance of the enumeration of the various (related) classes of polymers and ligation-cleavage reactions to the degree of catalysis required for the emergence of RAFs.
Recall that a CRS $\Q=(X, R, C, F)$ consists not only of a molecule set $X$ and the set of reactions $R$,  but also an assignment of catalysis $C$ and a food set $F$. 
To emphasise that $\Q$ depends on the maximal polymer length $n$, we will often write this as $\Q_n$.  
In polymer models, $F$ is typically taken to be all words of length at most  $t$ (where $t$ is typically small and independent of $n$ (e.g. $t=2$)). 
As for catalysis, this is assigned randomly, and various models have been proposed. The simplest (dating back to Kauffman ~\cite{kau86, kau93}), assumes that each molecule
catalyses each reaction with a fixed probability $p=p_n$ (possibly dependent on $n$) and  that such events are independent across all pairs $(x,r) \in X \times R$.

Thus each molecule type catalyses an expected number of $\mu_n = p_n |R|$ reactions.  Notice that if $p_n$ is independent of $n$, then $\mu_n$  grows exponentially with $n$; however, 
it turns out that $\mu_n$ needs only to grow linearly with $n$ for RAFs to arise with high probability. Moreover, there is a sharp transition here in the following sense (which following from results 
in \cite{mos05}). In a polymer model with oriented polymers and either under Convention A or B above, we have (for any $\epsilon>0$):

$$\mu_n = n^{1-\epsilon} \Longrightarrow \lim_{n \rightarrow \infty} \PP(\mbox {there exists a }  RAF \mbox{ for }  \Q_n) = 0;$$
$$\mu_n = n^{1+\epsilon} \Longrightarrow \lim_{n \rightarrow \infty} \PP(\mbox {there exists a }  RAF \mbox{ for } \Q_n) = 1.$$
This linear transition has been observed in  numerous simulation studies (first in \cite{hor04}). 
A more fine-grained analysis (also from \cite{mos05}) shows that if we write $\mu=\lambda n$ then, for any fixed $n$:
\begin{equation}
\label{ppeqs}
\PP(\mbox {there exists a }  RAF \mbox{ for }  \Q_n) \rightarrow  \begin{cases}
0, &  \mbox{ as $\lambda \rightarrow 0$},\\
1, & \mbox{ as $\lambda$ grows.}
\end{cases}
\end{equation}

The linear dependence of catalysis rate on $n$ in the transition from having no RAF to having a RAF in the oriented polymer setting is essentially because  $n$ is asymptotic to the ratio of the number of reactions divided by the number of molecule types (i.e. $|R_n|/|X_n|)$).  For  more general `polymer-like' systems (including non-oriented polymers), there is an analogue of Eqn.(\ref{ppeqs}) in Theorem 1 of \cite{smi14}, where again the ratio of reactions to molecule types plays a key role.   This is the main reason why it is important to have an asymptotic  measure of the size of these sets (as provided in Proposition~\ref{prol}).



\section{Complexity results for closed RAFs} 
\label{sec:closed}

For the rest of the paper, we no longer restrict ourselves to the world of polymer CRS systems; instead, our results apply to any CRS.

One important notion that ties the structural properties of RAFs to chemical realism is to require a RAF to be \textit{closed}. Formally, given a CRS $\Q = (X,R,C,F)$,  a RAF $R' \subseteq R$ is said to be {\em closed} if there is no reaction in $R - R'$ that has all its reactants and at least one catalyst in ${\rm cl}_{R'}(F)$. In other words, a closed RAF is one in which every reaction that \textit{can} happen \textit{will} happen.  For example, in Fig. \ref{fig3}, the RAFs $\{r_3\}$, $\{r_1, r_2\}$ and $\{r_1, r_2, r_3, r_4, r_5\}$ (the maxRAF) are the only closed RAFs present. Note that the maxRAF is always closed, if it exists.

Closed RAFs are of particular relevance to evolutionary theories of RAFs and early metabolic cycles: by finding the closed subRAFs of a particular RAF,  it may be possible to trace back its `ancestral' history; that is, a sequence of `stable' states that could have initially lead to the production of the RAF (see \cite{hor18}, \cite{ste18})). In a recent paper \cite{hor18}, a direct, formal relationship between closed RAFs and a field known as Chemical Organisation Theory was established (adding to earlier links (see \cite{ste13})). Using this connection, a new type of algorithm was developed to enumerate the set of all closed subRAFs existing within the maxRAF \cite{hor18}. Although the new algorithm had reasonable performance for RAFs of size $\leq 200$, it was not shown to run in polynomial time \cite{hor18}. It has remained an open question (posed in \cite{hor18} and  
\cite{ste18}) as to whether such a polynomial-time algorithm exists. 

In this section, we present our complexity results surrounding closed RAFs. We solve the open problem posed in \cite{hor18, ste18} by demonstrating the NP-completeness of finding closed subRAF. We also we show that finding closed `uninhibited' RAFs is not fixed-parameter tractable (FPT) in the number of inhibitions (unlike the case for non-closed RAFs \cite{hor12b}).

To motivate this setting, we consider first two questions without the `closure' constraint. 
Given a  CRS $\Q = (X,R,C,F)$ that has a RAF,   let $\hat{x} \in X$ be any molecule type.
There are then simple polynomial-time algorithms to determine answers to each of the following  questions. 

\noindent Does $\Q$ have:
\begin{itemize}
\item[(i)] a RAF $R'$ that does not produce $\hat{x}$?
\item[(ii)] a RAF that is a strict subset of ${\rm maxRAF}(\Q)$?
\end{itemize}

For Problem (i), let $R^*_{\hat{x}}$
be the set of reactions in ${\rm maxRAF}(\Q)$ that do not produce $\hat{x}$. The answer to (i) is `yes' if and only if
the CRS $(X, R^*_{\hat{x}}, C_{|X \times R^*_{\hat{x}}}, F)$ has a RAF. 

For Problem (ii),  let $R^*_r ={\rm max}(RAF(\Q))- \{r\}$. The the answer to (ii) is `yes'  if and only if
the CRS $(X, R^*_r, C_{|X \times R^*_r}, F)$ has a RAF for some $r \in {\rm maxRAF}(\Q)$.

However, if we modify these two questions so as to require the desired RAF to be closed, they become much more difficult, as we now explain.

We start with the following problem:
Given a CRS $\Q = (X, R, C, F)$, does $\Q$ contain a closed RAF $R'$ that does not produce a certain molecule type $\hat{x} \in X$?
More formally stated:

\begin{enumerate}[itemsep=0mm]
    \item[] \textbf{PROBLEM}: \textit{Forbidden-molecule closed RAF}
    \item[] \textbf{INSTANCE}: A CRS $\Q = (X,R,C,F)$ and a particular molecule type $\hat{x} \in X$.
    \item[] \textbf{QUESTION:} Does $\Q$ contain a closed RAF $R' \subseteq R$ with $\hat{x} \not\in
{\rm cl}_{R'}(F)?$
\end{enumerate}

\begin{Th}
\label{central} 
The \textit{Forbidden-molecule closed RAF} problem is NP-complete. 
\end{Th}
The proof Theorem~\ref{central} involves a delicate reduction from 3SAT (3-satisfiability), and is presented in the Appendix.

Theorem~\ref{central} provides the tool for establishing the hardness of some questions posed in earlier papers \cite{hor18, ste18}.

\begin{enumerate}[itemsep=0mm]
    \item[] \textbf{PROBLEM}: \textit{Closed strict subRAF}
    \item[] \textbf{INSTANCE}: A CRS $\Q = (X,R,C,F)$
    \item[] \textbf{QUESTION:} Does $\Q$ contain a closed RAF $R' \subset R$ with $R' \subsetneq \text{maxRAF}(\Q)$?
\end{enumerate}

\begin{Th}
The closed strict subRAF problem is NP-complete.
\end{Th}

{\em Proof:}   Given a set of reactions $R' \subseteq R$, one can check whether or not it  is both closed and is a strict subset of $\text{maxRAF}(\Q)$ in polynomial-time, so the problem is in the class NP. To establish NP-completeness, we perform a polynomial-time reduction from the Forbidden-molecule closed RAF problem (which is NP-complete by Theorem \ref{central}). Given a CRS $\Q = (X, R, C, F)$ and a particular molecule type $\hat{x} \in X$, we construct a CRS $\hat{\Q}= (\hat{X}, \hat{R}, \hat{C}, \hat{F})$ such that $\Q$ has a closed RAF $R' \subseteq R$ with $\hat{x} \not\in {\rm cl}_{R'}(F)$ if and only if $\hat{\Q}$ has a closed subRAF $R'' \subset \hat{R}$ that is a strict subset of the maxRAF in $\hat{\Q}$. 
The construction of $\hat{\Q}$ from $\Q$ is as follows. Let $\hat{\Q} = (\hat{X}, \hat{R}, \hat{C}, \hat{F})$   where $\hat{X} = X \cup \{\hat{f}\}$, $\hat{R} = R \cup \{\{\hat{f}\} \rightarrow \{\hat{x}\}\}$, $\hat{C} = C \cup \{ (x, r) : r \in \hat{R}\}$ and $F = F \cup \{\hat{f}\}$, for some newly introduced food molecule type $\hat{f}$. Essentially, we add a new food molecule type $\hat{f}$ to $\hat{\Q}$ and a reaction $\{\hat{f}\} \rightarrow \{\hat{x}\}$ that produces the forbidden molecule $\hat{x}$; we then let $\hat{x}$ catalyse every reaction in $\hat{\Q}$. This construction is clearly polynomial-time computable in the size of $\Q$.  We now note that since $X \subseteq \hat{X}$, $R \subseteq \hat{R}$, $C \subseteq \hat{C}$ and $F \subseteq \hat{F}$, we have $\text{maxRAF}(\Q) \subseteq \text{maxRAF}(\hat{\Q})$; furthermore, since $\hat{\Q}$ contains the additional reaction $\{\hat{f}\} \rightarrow \{\hat{x}\}$, we have $\text{maxRAF}(\Q) \subset \text{maxRAF}(\hat{\Q})$. 

We now establish the correctness of the construction. \\
\noindent ($\Longrightarrow$) Suppose that $\Q$ has a closed RAF $R' \subseteq R$ with $\hat{x} \not\in {\rm cl}_{R'}(F)$. Since $R'$ is a RAF in $\Q$, we have $R' \subseteq \text{maxRAF}(\Q)$. Now, as $\text{maxRAF}(\Q) \subset \text{maxRAF}(\hat{\Q})$, we have $R' \subset \text{maxRAF}(\hat{\Q})$. In other words, $R'$ is a strict subset of the maxRAF in $\hat{\Q}$. Furthermore, since $X \subseteq \hat{X}$, $R \subseteq \hat{R}$, $C \subseteq \hat{C}$ and $F \subseteq \hat{F}$, it follows that $R'$ is also a RAF in $\hat{\Q}$. We now complete the implication by showing that $R'$ is closed in $\hat{\Q}$. To do this, we first note that ${\rm cl}_{R'}(F) = {\rm cl}_{R'}(\hat{F})$ across $\Q$ and $\hat{\Q}$, respectively, for any $R' \subseteq  R$. This follows as for any $R' \subseteq R$ we have $\{\hat{f}\} \rightarrow \{\hat{x}\} \not\in R'$.
Now suppose that $r \in \hat{R} - R'$. There are two cases to consider: (i): $r \in R - R'$ and (ii)  $r = \{\hat{f}\} \rightarrow \{x\}$. 
In Case (i),  ${\rm cl}_{R'}(F)$ is closed in $\Q$, and as ${\rm cl}_{R'}(F) = {\rm cl}_{R'}(\hat{F})$, we have $R'$ closed in $\hat{\Q}$. In Case (ii),  since  $R' \subseteq R$, we have  $\{\hat{f}\} \rightarrow \{\hat{x}\} \not\in R'$. With this, and since $\hat{x} \not\in {\rm cl}_{R'}({F})$ in $\Q$, it follows $\hat{x} \not\in {\rm cl}_{R'}(\hat{F})$ in $\hat{\Q}$ and so $r = \{\hat{f}\} \rightarrow \{\hat{x}\}$  is uncatalysed across ${\rm cl}_{R'}(\hat{F})$. Therefore $R'$ is closed in $\hat{\Q}$, and thus  $R'$ is a closed, strict subRAF of the maxRAF of $\hat{\Q}$. 

\noindent $(\Longleftarrow$) Suppose that $R'' \subset \hat{R}$ is a closed RAF in $\hat{\Q}$ that is a strict subRAF of $\text{maxRAF}(\hat{\Q})$. Since $R'' \subset \text{maxRAF}(\hat	{\Q})$, we must have $\hat{x} \not\in {\rm cl}_{R''}(\hat{F})$ (otherwise $\hat{x}$ would catalyse every reaction in $\hat{\Q}$, and so,  by the closure of $R''$, it could no longer be a strict subset of $\text{maxRAF}(\hat{\Q})$). With $\hat{x} \not\in {\rm cl}_{R''}(\hat{F})$,  we have $\{\hat{f}\} \rightarrow \{x\} \not\in R''$, and therefore $R'' \subseteq R$ in $\Q$. Furthermore, as ${\rm cl}_{R'}(F) = {\rm cl}_{R'}(\hat{F})$ across $\Q$ and $\hat{\Q}$ for any $R' \subseteq  R$  (as mentioned previously), and by construction,  $R''$ must also be a RAF in $\Q$. Since $X \subseteq \hat{X}$, $R \subseteq \hat{R}$, $C \subseteq \hat{C}$ and $F \subseteq \hat{F}$ (and as $R''$ is closed in $\hat{\Q}$), $R''$ must also be closed in $\Q$ with $\hat{x} \not\in {\rm cl}_{R'}(F)$. This completes the argument.

\hfill$\Box$

\subsection{Application to `uninhibited' RAFs}
\label{urafsec}

So far, the CRS model only includes catalytic interactions between molecule types and reactions. This may be plausible in some circumstances, although in real biochemical systems, it is very often the case that certain molecule types \textit{inhibit} the presence of certain reactions. 
To describe this more formally, consider a CRS $\Q = (X, R, C, F)$ together with an additional set $I \subseteq X \times R$. An {\em uninhibited RAF} (uRAF) is a RAF set $R'$ for $\Q$ that satisfies the following additional property: for every reaction $r \in R'$,  no molecule type in the food set or produced by another reaction in $R'$ inhibits $r$.    Formally: For each $ r \in R'$, there is no molecule type $x \in {\rm cl}_{R'}(F)$ for which $(x, r) \in I$ \cite{hor15}. 
Essentially, inhibition can be viewed as the exact opposite of catalysation.  Note that a reaction that is inhibited can still be part of a RAF, providing that the only  molecule types that inhibit the reaction
are not present in ${\rm cl}_{R'}(F)$. 

Following \cite{smi14}), it seems sensible to further restrict interest to uRAFs that are also closed RAFs, since an uninhibited \textit{non-closed} RAF could provide the reactants and catalyst for one or more additional reactions (outside the RAF) to occur, which, in turn, generate a a molecule that inhibits a reaction within the original RAF. 
 Closed uRAFs, on the other hand, are truly free of inhibition from products of such reactions. It was shown in \cite{mos05} that determining the existence of a (non-closed) uRAF within a CRS $\Q$ is NP-hard, though it was shown in  \cite{hor12b} to be fixed-parameter-tractable in the number of inhibiting molecule types.

For non-closed RAFs,  an algorithm exists for finding uRAFs that is fixed-parameter-tractable in the number of inhibiting molecule types \cite{hor12b}. We prove the same result cannot be found for closed uRAFs (subject to $P \neq NP$) by showing that the following question is NP-complete.
\begin{Th}
The following problem is NP-complete. Given a CRS $\Q = (X, R, C, F)$ and a \textit{single} inhibiting pair $(x, r) \in X \times R$, determine whether or not a closed uRAF for $(\Q, I)$ exists.\end{Th}
{\em Proof:}
Checking that a set $R'$ is a closed uRAF is decidable in polynomial-time, so the problem is in the class NP.
To establish NP-completeness, we again perform a polynomial-time reduction from the Forbidden-molecule closed RAF problem ({\em c.f.} Theorem~\ref{central}).  Given a CRS $\Q = (X, R, C, F)$ and a particular molecule type $\hat{x} \in X$, we construct, in polynomial time, a CRS $\Q'$ and an inhibiting set $I \in X \times R$ such that $\Q$ has a closed RAF $R' \subseteq R$ with $\hat{x} \not\in {\rm cl}_{R'}(F)$ if and only if $\Q'$ has a closed, uninhibited RAF. The construction is simple: let $\Q' = Q$ and define $I = \{(x, r) : r \in R\}$ (i.e.  make $x$ inhibit all reactions in $R$). 
It follows that  a closed RAF $R'$ in $\Q$ with $\hat{x} \not\in {\rm cl}_{R'}(F)$ will be a closed, uninhibited RAF in $\Q'$. 
Conversely, a closed uRAF $R'$  for $\Q'$ is a closed RAF for which $\hat{x} \not\in {\rm cl}_{R'}(F)$.  Moreover, the  construction is clearly polynomial-time computable, as required. 
\hfill$\Box$

\section{Complexity results for elementary RAFs} 
\label{sec:compl}

To help avoid NP-complete problems arising in RAF theory, a simpler setting has recently been studied (see \cite{ste18}).  An {\em elementary} CRS is a special type of CRS where each reaction has all its reactants in the food set. Every reaction is therefore trivially F-generated and so the RAF condition reduces to just the  RA catalysis condition (i.e. each reaction is catalysed by the product of some other reaction or by an element of the food set).

Although this setting seems quite restrictive, it is nevertheless pertinent in experimental systems (\cite{vai12}, \cite{ash04}) as well as theoretical models (\cite{jai98}). Many problems that are known to NP-hard in the general setting are tractable in the elementary setting: for example, finding the minimum-size RAF  and finding closed subRAFs of the maxRAF (proven to be NP-hard in Section~\ref{sec:closed}) are both computable in polynomial-time in the elementary setting \cite{ste18}. We now present two  questions  concerning  elementary CRSs that were posed recently in \cite{ste18}:
\begin{itemize}
\item[{\bf Q1}]  {Is there a polynomial-time algorithm to find a uRAF for an elementary CRS?}
\item[{\bf Q2}] {Is there a polynomial-time algorithm to find a {\em maximum}-sized irreducible RAF  for an elementary CRS?}

\end{itemize}

Here, we resolve both questions Q1  and Q2 by providing proofs of NP-hardness. 

First recall that a uRAF is a RAF set $R' \subseteq R$ that satisfies the additional property: for each reaction $r \in R'$, an inhibiting molecule type for $r$ is not present in either the food set or as a product of another reaction in $R'$. \cite{hor15}.   
\begin{Th}
\label{elthm}
Given an elementary CRS $\Q = (X, R, C, F)$ and an inhibition assignment $I \subseteq X \times R$, determining whether or not a uRAF for $\Q$ exists is NP-complete.
\end{Th}
The proof of this result is provided in the Appendix.

\bigskip

We turn now to Question Q2.   When a RAF set contains no subRAFs, it is said to be an \textit{irreducible} RAF or, more briefly, an {\em  irrRAF}.  Note that a RAF is an irrRAF if it has the property that removing any single reaction from it results in a set that contains no RAF.  For example, in  Fig. \ref{fig3}, the sets $\{r_3\}$ and $\{r_1, r_2\}$ are the only irrRAFs present.  Irreducible RAFs have been extensively used in structural analyses of RAFs in polymer models \cite{hor13, ste13} and in extant metabolic systems,  \cite{sou15} and to model  `coherent evolutionary units',  an evolutionary analogue of cells  (see \cite{vas12}). 
Finding a single irrRAF within a RAF is easily shown to be  computable in polynomial-time. However, finding a smallest irrRAF (or, equivalently, a minimum-size RAF) turns out to be  NP-hard  \cite{	hor13}.

It has been shown earlier \cite{ste13} that the problem of determining the size of a smallest RAF (which is necessarily an irrRAF) in a CRS is NP-hard.  However, the problem of  determining the size of a largest irrRAF in a CRS was previously of unknown complexity, and was a problem posed in \cite{sou15}.  Here, we show that this problem is not only NP-hard but it remains NP-hard when we restrict it to the setting of elementary CRS systems, which, in turn, answers a question posed in 
 \cite{ste18}.

\begin{Th}
\label{irrRAFth}
Determining the size of a largest  irrRAF in an elementary CRS $\Q$ is NP-hard.
\end{Th}

\begin{figure}[ht]
\begin{center}
\includegraphics[width=0.5\linewidth]{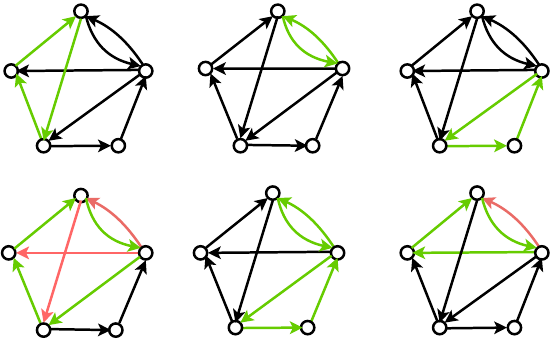} 
  \caption{The top three cycles (paths in green) are simple chordless cycles. 
  The bottom three cycles (chords highlighted in red) are not simple chordless cycles (the bottom-middle one is not simple; for the other two,  chords are highlighted in red).}
    \label{fig5}
\end{center}
\vspace{-12pt}
\end{figure}

To establish this result, we need to introduce the longest directed simple chordless cycle problem (see also the `snake-in-the-box' problem \cite{kau58}), which is known to be NP-hard \cite{gar90}.
For a digraph $G = (V, E)$, a simple chordless cycle is a sequence of distinct vertices $v_0, v_1,\ldots, v_{k-1}$ with $(v_{i}, v_{(i+1) \text{ mod } k}) \in E$ for each $i < k$, and where, for every pair of vertices $v_i, v_j$ for $j \neq (i+1) \text{ mod } k$, we have no edge $(v_i, v_j) \in E$. See Fig. \ref{fig5} for an illustration.

\begin{figure}[ht]
\begin{center}
\includegraphics[width=0.5\linewidth]{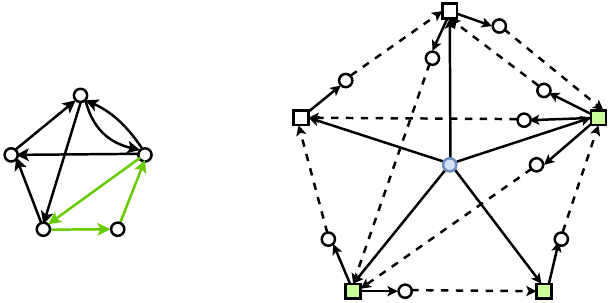} 
  \caption{A digraph $G$ displayed on the left, with the constructed elementary CRS $\Q_G$ on the right (the food molecule is at the centre, highlighted in blue). The  simple chordless cycle in $G$ of size 3 (highlighted in green) corresponds to an irrRAF in $\Q_G$ of size 3.}
    \label{fig6}
\end{center}
\vspace{-20pt}
\end{figure}

Given a digraph $G = (V, E)$, we first describe a simple polynomial-time construction of a certain elementary CRS $\Q_G = (X, R, C, F)$ as follows: for each $v \in V$, let $v^R \in R$ be a reaction in $\Q_G$; for each $(u, v) \in E$, let $u_v^X \in X$ be a molecule produced by $u^R$ that catalyses the reaction $v^R$ in $\Q_G$; let $F = \{f\}$ with $f$ being the sole reactant of all reactions.  Since we introduce a linear number of reactions, molecule types and edges with respect to $|G|$, our construction is computable in polynomial-time. See Fig. {\ref{fig6}} for an illustration.

Since every reaction is trivially F-generated, and since $f$ does not catalyse any reactions, we can simplify the RAF definition in this setting: a set $R' \subseteq R$ will be a RAF set in $\Q_{G}$ if and only if each reaction $r \in R'$ is catalysed by a molecule which is produced by another reaction $r' \in R'$. 
It follows that a set of reactions $R' = \{v_0^R, v_1^R, \ldots, v_{k-1}^R\}$ is a RAF in $\Q_G$ if and only if for each $v_i^R \in R'$, there is a reaction $v_j^R \in R'$ that produces the molecule $v_j^X \in {\rm cl}_{R'}(F)$ with $(v_j^X, v_i^R) \in C$.  By construction, this happens if and only if $(v_j, v_i) \in E$. 

Furthermore, let us now consider a set  $V' = \{v_0, v_1,\ldots, v_{k-1}\} \subseteq V$ of distinct vertices in $G$ such that for each $v_i \in V'$ there exists a $v_j \in V'$ with $\{v_j, v_i\}$ in $E$.  By the construction of $\Q_G$, it follows that  $R' = \{v_0^R, v_1^R, \ldots, v_{k-1}^R\}$ will have every reaction $v_i^R \in R'$ catalysed by a molecule $v_j^X \in {\rm cl}_{R'}(F)$ produced by $v_j^R \in R'$, and therefore $R'$ is a RAF set. 

Combining these together, $R' = \{v_0^R, v_1^R, \ldots, v_{k-1}^R\}$ is a RAF in $\Q_G$ if and only if for every vertex $v_i \in V' = \{v_0, v_1,\ldots, v_{k-1}\} \subseteq V$, there exists another vertex $v_j \in V'$ with $(v_j, v_i) \in E$. Formally, fix a  set $V' = \{v_0, v_1, \ldots,  v_{k-1}\} \subseteq V$, then:
\begin{equation}
\label{disjointcy}
\forall v_i \in V' , \exists v_j \in V' :  (v_j, v_i) \in E  \Longleftrightarrow \{v_0^R, v_1^R, \ldots,  v_{k-1}^R\} \subseteq R \text{ is a RAF in $\Q_G$.}
\end{equation}

Recall the definition of an irrRAF: a RAF set $R'$ is said to be irreducible (irrRAF) if there is $R'$ contains no RAF as a strict subset.
The following lemma links irrRAFs to simple chordless cycles in $\Q_G$.
\begin{lem}
\label{niceIrraf}
Fix a digraph $G = (V, E)$ and a number $k \geq 1$. Then  $v_0, \ldots,  v_{k-1}$ is a simple chordless cycle in $G$ if and only if $\{v_0^R, \ldots, v_{k-1}^R\}$ is an irrRAF in $\Q_G$.
\end{lem}

{\em Proof:}

($\implies$) Suppose that $v_0, \ldots, v_{k-1}$ is a simple chordless cycle in $G$ for $k \geq 2$. Since $v_0,\ldots,v_{k-1}$  is simple, we can define a set $V' = \{v_0,\ldots,v_{k-1}\} \subseteq V$ (simple implying no deduplication when translating to a set). As $v_0,\ldots,v_{k-1}$ is a cycle, the  LHS of (\ref{disjointcy}) is true, so $R' = \{v_0^R, v_1^R, \ldots, v_{k-1}^R\} \subseteq R$ is a RAF in $\Q_G$.  We now show that $R'=  \{v_0^R, v_1^R, \ldots, v_{k-1}^R\}$ is irreducible. Since   $v_0,\ldots,v_{k-1}$ is chordless in $G$, it follows that for all $0 \leq i < k$, the only vertex inbound to $v_i$ from other vertices in $V'$ is $v_{(i-1) \textit{ mod } k}$. By construction, we therefore see that for each reaction $v_i^R \in R'$ we have $v_{(i-1) \text {mod } k}^R$ the \textit{only} reaction in $R'$ producing a molecule that catalyses $v_i^R \in R'$. It follows that no strict subset of reactions $R'' \subset R$ is a RAF; otherwise, for some $0 \leq j < k$ we would have $v_{(j-1) \text {mod } k}^R \not\in R''$ with $v_j^R \in R''$ and $v_j^R$ not being catalysed. Therefore $R' = \{v_0^R, v_1^R ,\ldots ,v_{k-1}^R\}$ is an irrRAF in $\Q_G$.

($\Longleftarrow$) Suppose that $R' = \{v_0^R, v_1^R ,\ldots ,v_{k-1}^R\}$ is an irrRAF in $\Q_G$. Let $V' = \{v_0, v_1,\ldots, v_{k-1}\} \subseteq V$ be a distinct set of vertices from $G$. As $R'$ is a RAF by (\ref{disjointcy}) it follows that for every vertex $v_i \in V'$ there exists another other vertex $v_j \in V'$ with $(v_j, v_i) \in E$ (i.e.  $\forall v_i \in V',  \exists v_j \in V' : (v_j, v_i) \in E$). If we now consider any set $V'' \subset V'$, the same cannot apply else by (\ref{disjointcy}) we would see $V''$ corresponding to a subRAF $R'' \subseteq R'$. As $R'$ is irreducible, this is a contradiction. It follows that for every vertex $v_i \in V'$, we have $(v_j, v_i) \in E$ for exactly one vertex $v_j \in V'$.  The set $V'$ must form a simple chordless cycle. 
\hfill$\Box$

\bigskip

{\em Proof of Theorem~\ref{irrRAFth}:}
We apply Lemma \ref{niceIrraf} to the complexity questions surrounding RAF irreducibility. Recall that this lemma states that 
if we fix a digraph $G = (V, E)$ and a number $k \geq 2$, then $v_0,\ldots, v_{k-1}$ is a simple chordless cycle in $G$ if and only if $\{v_0^R,\ldots, v_{k-1}^R\}$ is an irrRAF in $\Q_G$.
Finding the longest simple chordless cycle is a known NP-hard problem \cite{gar90}, so the proof follows immediately from Lemma \ref{niceIrraf}.
\hfill$\Box$


\section{Concluding comments}
\label{sec:conc}

In this paper, we have
provided the first exact formula for the number of reactions in polymer models, and established asymptotic results from them that verify earlier results that were based on heuristic 
approximations.  We have also settled a number of computational complexity questions concerning the detection of subRAFs of particular types. 
As many of the interesting questions turn out to be NP-hard, it may be helpful to develop tailored SAT-solver or Linear Programming techniques  for analysing CRS datasets (e.g. to determine uRAF and minimum RAF sizes), and to thus extend the computational results beyond what was previously tractable. We plan to investigate these in future work.

\section{References}

\label{Bibliography}


\bibliographystyle{siamplain}
\bibliography{RAF}

\section{Appendix: Proofs  of Theorems~\ref{main1}(i), ~\ref{central} and ~\ref{elthm}.}

\bigskip


\noindent {\bf Proof of Theorem~\ref{main1} Part (i):}

{\em Part (i):}   The proof consists of establishing a series of claims. 

 \noindent {\bf Claim 1:} Using Convention B to count $U(i)$ results  in double-counts precisely whenever $uv = vu = z, u \neq v$ and $|z| = i$. 
 
To see this, observe that for $u,v,z \in \Sigma^+$ with $uv = z$, we double-count using Convention B (instead of Convention A) precisely when there are two distinct triples $(u,v,z)$ and 
$(u', v', z')$ with $uv = u'v' = z$ and the corresponding sets $\{u, v, z\} = \{u', v', z\}$ equivalent. Since $z$ is common to both sets, we have $\{u, v, z\} = \{u', v', z\}$ if and only if $\{u, v \} = \{u', v' \}$; however, since $(u, v, z)$ and $(u', v', z)$ are distinct, we also have $u \neq u'$ or $v \neq v'$. This gives us $\{u,v,z\} = \{u', v',z\}$ if and only if $u = v'$ and $v = u'$. Rearranging gives $uv = vu = z$. This establishes  Claim 1.

\bigskip

 \noindent {\bf Claim 2:} For each fixed $i \geq 2$ there is a bijection between the set of periodic words $w_i$ of length $i$ and the set of aperiodic sub-words $w_d$ of length $d$ where $d | i$ and $d < i$. 

To see this, observe that every periodic word $w_i$ must have a unique period, as the period is the \textit{smallest} sub-word that repeatedly joins to form the word; further, that period is an aperiodic sub-word of length $d < i$ with $d | i$. Next, fix an aperiodic word $w_d$ of length $d$ with $d | i, d < i$; then $w_d^{i / d} = w_i$, so $w_d$ uniquely corresponds (as a sub-word) to a periodic word $w_i$ of length $i$ with $d | i , d < i$. 
This establishes Claim 2.

With this bijection, we can easily derive a recursive formula to count the number of aperiodic words of length $i$.

\bigskip

 \noindent {\bf Claim 3:} The number of aperiodic  words $\ovl{p}(i)$ of length $i \geq 2$ is given exactly by:
\begin{equation} \label{ap}
    \ovl{p}(i) = k^i - \sum_{d|i, d < i} {\ovl{p}(d)}
\end{equation}

To see this, observe that  the number of aperiodic words of length $i$ is clearly $k^i$ subtract the number of periodic words of length $i$. By Claim 2, we can count the number of periodic words of length $i$ by counting the number of aperiodic words of length $d < i, d | i$. This gives Eqn.(\ref{ap}), and so establishes Claim 3.

We make a small point about the base case: technically, $\ovl{p}(1)$ is not defined above (a single character is aperiodic) though we still use the value of $\ovl{p}(1) = k$ as the base-case for $i \geq 2$. With this, we now achieve a non-recursive formula. Rearranging (\ref{ap}), we get:
\begin{equation} \label{map}
    {k}^i = \sum_{d|i}{\ovl{p}(d)}
\end{equation}
By applying the M\"obius inversion  formula we get:
\begin{equation}
        \ovl{p}(i) = \sum_{d | i}{\mu\left(\frac{d}{d'}\right)k^{d'}}
\end{equation}
Recalling our original problem, where we double-count precisely whenever $uv = vu = z$, $u \neq v$ and $|z| = i$ for any $u,v,z \in \Sigma^+$, it follows from Lemma \ref{abba} that $z$ must therefore be periodic. 

\bigskip

\noindent {\bf Claim 4:} Using Convention B (rather than A), the number of double-counts per periodic word of length $i$ with period length $d$ is given exactly by $\left \lfloor \frac{i/d -1}{2} \right \rfloor$.

To see this, let  $w_i$ be a periodic word of length $i$ and $w_d$ its period of length $d$. It follows that $w_{d}^jw_{d}^{i/d-j} = w_i$ for each $1 \leq j < i/d$. Using Convention B, we will double-count precisely whenever $j < i/d - j$, as $j$ is symmetric for $i/d-j$. For an example, let $w_i=abcabcabcabc$. Now,  $w_i$ has a period $abc$ and there are $3$ ways to partition $w_i$ with units $abc$ using Convention B:\begin{center}
    (1) $\{abc,   {abc}{abc}{abc},   w_i\}$
    \item (2) $\{abcabc,  {abc}{abc},  w_i\}$
    \item (3) $\{abcabc{abc},  {abc},  w_i\}$
\end{center}

However,  (1) and (3) are the same sets --- and therefore double-counted in the unordered version. Combining $j < i/d - j$ with $j \geq 1$ gives us $\left\lfloor\frac{i/d-1}{2} \right\rfloor$ double-counts. This establishes Claim 4.

Next, recall that $U(i)$ denote the number of unique sets $\{u,v,z\}$ for $u,v,z \in \Sigma^+$ such that $uv = z$ and $|z| = i$. 

\bigskip

\noindent{\bf Claim 5:}  $U(i)$ is given by the expression in Eqn.~(\ref{rMob}).

To establish  Claim 5 we use Convention B (which leads to Eqn.~(\ref{eqs})) to give  $(i-1){k}^i$ and then subtract the double-counts
to give our desired Convention A.  From Claim 4, each period word of length $i$ and period $d$ will incur exactly ${\left \lfloor \frac{i/d - 1}{2}\right \rfloor}$ double-counts; further, from Lemma \ref{abba} we double-count exclusively when this is the case. This gives us the formula:
$${   U(i) = 
    (i-1){k}^n - \sum_{d|i, d<n}
    {
      {\left \lfloor \frac{i/d - 1}{2}\right\rfloor}
      \ovl{p}(d)
  }
}$$
Using the explicit formula for $\ovl{p}(d)$, we arrive at Eqn. (\ref{rMob}).
This establishes Claim 5, and thereby Part (i). 

\hfill$\Box$

\bigskip

{\bf Proof of Theorem~\ref{central}:}  Our proof will follow a reduction from 3SAT. Given a formula $\mathcal{F}$, we will construct in polynomial time a CRS $\Q_{\mathcal{F}} = (X,R,C,F)$ with a distinguished molecule $\hat{x} \in X$ that has a closed RAF $R'$ with $\hat{x} \not\in {\rm cl}_{R'}(F)$ if and only if $\mathcal{F}$ admits a satisfying assignment.

\begin{figure}[ht]
\begin{center}
\includegraphics[width=0.6\linewidth]{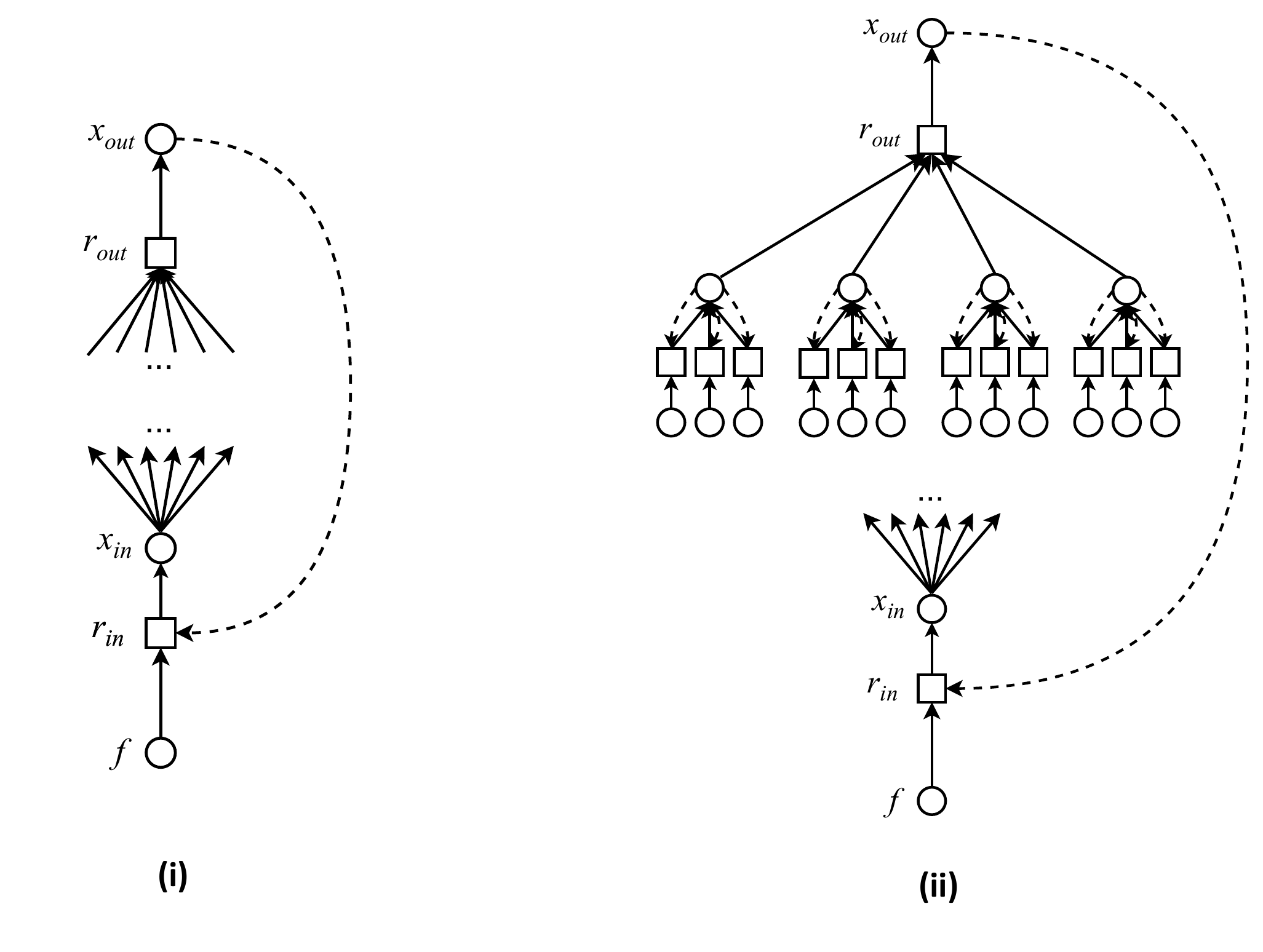} 
  \caption{(i) The food gadget (incomplete):  both $r_{out}$ and $r_{in}$ must necessarily be in any RAF set of $\Q_{\mathcal{F}}$. (ii) The clause gadget: at least one reaction from each `block' (of three) in the middle row must be in any RAF set $R' \subseteq R$.}
    \label{fig7}
\end{center}
\end{figure}

We start by describing the \textit{food gadget}.  Our CRS $\Q_{\mathcal{F}}$ will contain a single food molecule $F = \{f\}$. We will include in $\Q_{\mathcal{F}}$ two distinguished molecule types $x_{out}$ and $x_{in}$ as well as two distinguished reactions $r_{\text{out}}$ and $r_{\text{in}}$. We present the food gadget in Fig.~\ref{fig7}(i) (for now we exclude the reactants of $r_{out}$ and the reactions that $x_{in}$ acts as a reactant for).

Since $r_{in}$ is (and will be, in the overall construction) the only reaction with all food-molecule reactants,  any F-generated set in $\Q_{\mathcal{F}}$ must contain $r_{in}$ (in fact, any linear ordering of reactions must start with $r_{in}$). As RAFs are by definition F-generated, the following condition can therefore be inferred: 
\begin{equation}
\label{rin}
R' \subseteq R \text { is a RAF} \implies r_{in} \in R' 
\end{equation}
As $x_{out}$ is the only molecule catalyzing $r_{in} \in R'$ and $r_{out}$ is the only reaction producing $x_{out}$, we must also have:
\begin{equation}
\label{rout}
R' \subseteq R \text { is a RAF} \implies r_{out} \in R'. 
\end{equation}
Next, we present the \textit{clause gadget}, displayed in Fig. \ref{fig7}(ii).
From (\ref{rout}), we can infer that each reactant of $r_{out}$ must also be in ${\rm cl}_{R'}(F)$. Let $A_{out} \subseteq X$ denote the set of reactants of $r_{out}$. It follows that if $R' \subseteq R$ is a RAF then $\bigwedge_{x \in A_{out}}{x \in {\rm cl}_{R'}(F)}$. 

Further, since none of $x \in A_{out}$ are food molecule types, we must have at least one reaction producing $x$. Let $\rho: X \rightarrow 2^R$ denote the set of reactions that produce a molecule $x \in X$. The following holds: \begin{equation}
\label{irrP1}
R' \subseteq R \text { is a RAF} \implies \bigwedge_{x \in A_{out}}{\bigvee_{r \in \rho(x)}{r \in R'}} 
\end{equation}
In Fig. \ref{fig7}(ii), this condition forces at least one reaction from each `block' (of three reactions) in the middle row to be in $R'$ --- for \textit{any} RAF set $R' \subseteq R$. 

We will now integrate our clause gadgets with the structure of the given formula $\mathcal{F}$. For each clause $c_{i} = l_{i1} \wedge l_{i2} \wedge l_{i3} \in \mathcal{F} = \bigwedge_{i} c_i$, 
let $c_i^X$ be a reactant molecule to $r_{out}$ (i.e. let $c_i^X \in A_{out}$) and let $l_{i1}^R, l_{i2}^R, l_{i3}^R$ be reactions producing $c_i^X$  (i.e. let $l_{i1}^R, l_{i2}^R, l_{i3}^R \in \rho(c_i^X)$). For each reaction $l_{ij}^R$, also let $l_{ij}^X$ be its single reactant and let $c_i^X$ be its single catalysing molecule. Stated formally: for each clause $c_i \in \mathcal{F}$, we have $l_{ij} \in c_i \implies (l_{ij}^R = (\{l_{ij}^X\}, \{c_i^X\}) \in R) \wedge ((c_i^X, l_{ij}^R) \in C)$ for $1 \leq j \leq 3$.

We can now rewrite condition (\ref{irrP1}) as:
\begin{equation}
\label{irrP2}
R' \subseteq R \text { is a RAF} \implies \bigwedge_{i}{(l_{i1}^R \in R') \vee (l_{i2}^R \in R') \vee (l_{i3}^R \in R')} 
\end{equation}
Since each $l_{ij}^R$ reaction from \ref{irrP2} must contain its reactants we also gain the condition: 
\begin{equation}
\label{irrP3}
R' \subseteq R \text { is a RAF} \implies \bigwedge_{i}{(l_{i1}^X \in {\rm cl}_{R'}(F)) \vee (l_{i2}^X \in {\rm cl}_{R'}(F)) \vee (l_{i3}^X \in {\rm cl}_{R'}(F))} 
\end{equation}
As we proceed, will refer to $l_{ij}^R \in R$ as the `literal' reaction and $l_{ij}^X$ as the `literal' molecule for the literal occurrence $l_{ij} \in \mathcal{F}$.

\begin{figure}[ht]
\begin{center}
\includegraphics[width=0.3\linewidth]{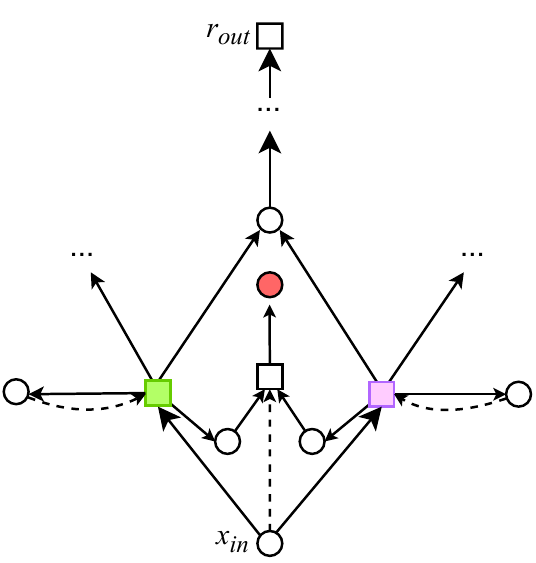} 
  \caption{The variable gadget for a variable $v \in \mathcal{F}$: in any closed RAF set $R' \subseteq R$ that does not produce $\hat{x}$ we have exactly one of $v^R, \ovl{v}^R$ in $R'$.}
    \label{fig8}
\end{center}
\end{figure}

Next, we describe the \textit{variable gadget}, displayed in Fig. \ref{fig8}.
For each variable $v \in \mathcal{F}$, introduce a variable gadget to $\Q_{\mathcal{F}}$. The variable gadget  is parametrised by a variable $v$ and consists of three reactions. It is displayed in Fig. \ref{fig8}. For a variable $v$, the variable gadget contains reactions named $v^R$ (highlighted in green), $\ovl{v}^R$ (highlighted in pink) and $v_{mid}^R$ (non highlighted: produces the red molecule). The central red molecule is $\hat{x} \in X$, which is the forbidden molecule. 
Formally, we have $v^R = (\{ x_{in} \}, \{ v_{cat}^X, v_{top}^X, v_{mid}^X\})$,  $\ovl{v}^R = (\{ x_{in} \}, \{ \ovl{v}_{cat}^X, v_{top}^X, \ovl{v}_{mid}^X\})$, $v_{mid}^R = (\{v_{mid}^X,  \ovl{v}_{mid}^X\}, \{\hat{x}\})$ for molecule types $v_{top}^X, v_{mid}^X, \ovl{v}_{mid}^X, v_{cat}^X, \ovl{v}_{cat}^X$ with $(v_{cat}^X, v^R), (\ovl{v}_{cat}^X, \ovl{v}^R) \in C$.
 
Since $v_{top}^X$ is a reactant to $r_{out}$, by (\ref{rout}) it follows that $v_{top}^X \in {\rm cl}_{R'}(F)$ for any RAF set $R' \subseteq R$; further, as $v_{top}^X$ is only produced by $v^R$ and $\ovl{v}^R$, we can infer the following:
\begin{equation}
\label{got1var}
R' \subseteq R \text{ is a RAF } \implies (v^R \in R') \vee (\ovl{v}^R \in R')
\end{equation}

Next, for any closed RAF $R' \subseteq R$, if both $v^R, \ovl{v}^R \in R'$, it follows $v_{mid}^X, \ovl{v}_{mid}, x_{in} \in {\rm cl}_{R'}(F)$ and therefore $v_{mid}^R \in R'$ (by closure of $R'$). Since $v_{mid}^R \in R'$ produces our forbidden molecule, we can infer (contrapositive) the following condition:
\begin{equation}
\label{gotleq2var}
R' \subseteq R \text{ is a closed RAF } \wedge \hat{x} \not\in {\rm cl}_{R'}(F)  \implies (v^R \notin R') \vee (\ovl{v}^R \notin R')
\end{equation}
Combining (\ref{gotleq2var}) with (\ref{got1var}) gives:
\begin{equation}
\label{varCond}
R' \subseteq R \text{ is a closed RAF } \wedge \hat{x} \not\in {\rm cl}_{R'}(F) \implies (v^R \in R') \Leftrightarrow (\ovl{v}^R \notin R')
\end{equation}

We will now consolidate our variable gadgets with our clause gadgets.

\begin{figure}[ht]
\begin{center}
\includegraphics[width=1.0\linewidth]{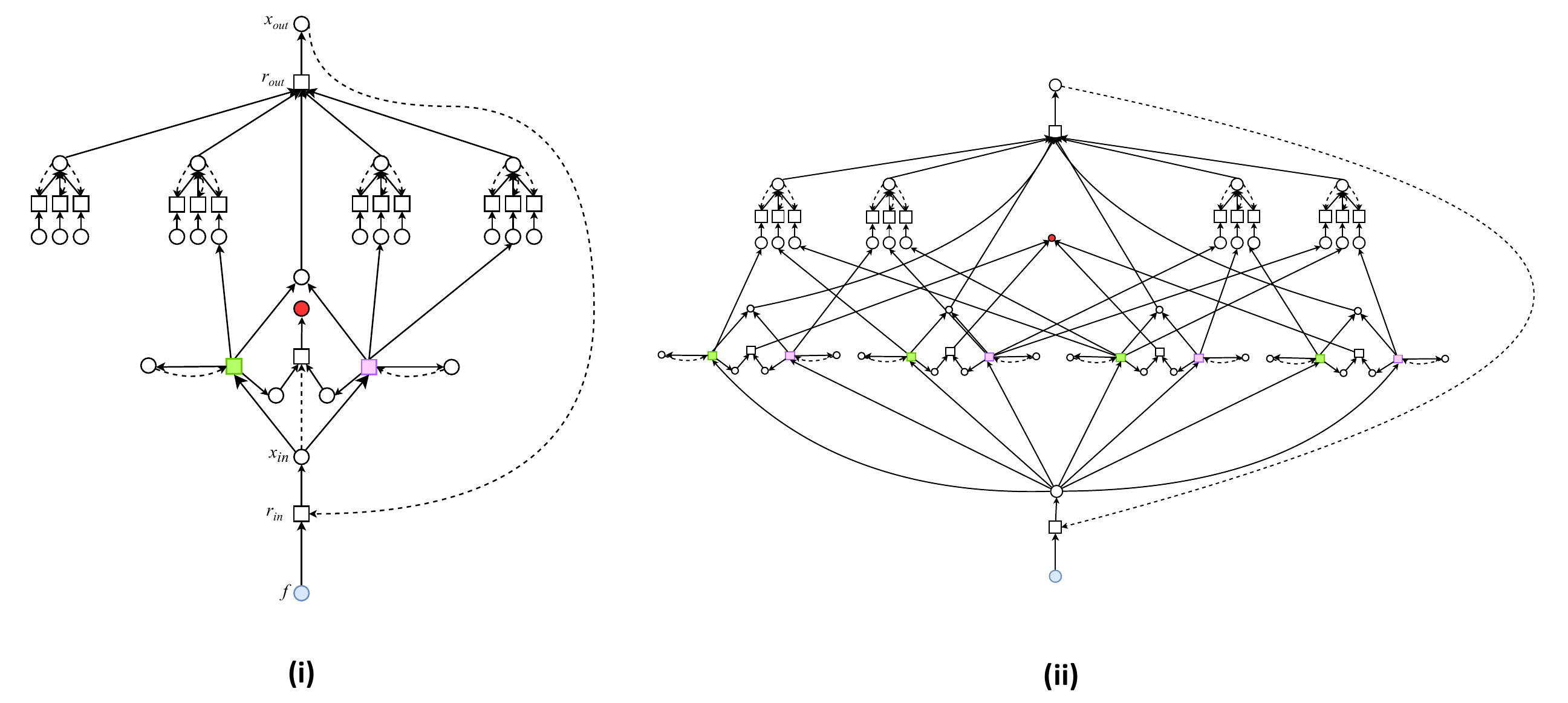} 
  \caption{(i) Part of the construction for the Formula $\mathcal{F} = (a \vee b \vee c) \wedge (a \vee \ovl{b} \vee x) \wedge (a \vee \ovl{x} \vee c) \wedge (a \vee \ovl{x} \vee d)$, only including the variable gadget for $x$ (hiding the gadgets for variables $a,b, c, d$). The food molecule is coloured blue. (ii) The entire construction $\Q_{\mathcal{F}}$ for the formula $\mathcal{F} = (a \vee b \vee c) \wedge (\ovl{a} \vee \ovl{b} \vee c) \wedge (\ovl{b} \vee \ovl{c} \vee d) \wedge (b \vee c \vee \ovl{d})$. The food molecule is coloured blue.}
    \label{fig9}
\end{center}
\end{figure}

For a variable gadget in $\Q_{\mathcal{F}}$ corresponding to a variable $v \in \mathcal{F}$, let $v^R$ produce all the literal molecule types $l_{ij}^X$ for $l_{ij} \in \mathcal{F}$ being a \textit{positive} occurrence of the variable $v$ --- i.e. when $l_{ij} = v \in c_i$. Likewise, let $\ovl{v}^R$ produce all literal molecule types $l_{ij}^X$ for $l_{ij} \in \mathcal{F}$ being a \textit{negative} occurrence of the variable $v$ --- i.e. when $l_{ij} = \ovl{v} \in c_i$. Each literal molecule $l_{ij}^X$ will be produced only by its corresponding variable gadget reaction; that is, either $\rho(l_{ij}^X) = \{v^R\}$ or $\rho(l_{ij}^X) = \{\ovl{v}^R\}$ for $l_{ij} = v$ or $l_{ij} = \ovl{v}$, respectively. Since these are singleton sets, we define $\rho_l$ to be the `unpacking' of the set --- i.e. $\rho_l(l_{ij}^X) = v^R$ or $\rho_l(l_{ij}^X) = \ovl{v}^R$. With this, we can rewrite (\ref{irrP3}) as: 
\begin{equation}
\label{irrP4}
R' \subseteq R \text { is a RAF } \implies \bigwedge_{i}{(\rho_{l}(l_{i1}^X) \in R') \vee (\rho_l(l_{i2}^X) \in R') \vee (\rho_l(l_{i3}^X) \in R')}
\end{equation}
This enables us to condition the structure of $\mathcal{F}$ over the variable gadgets in $\Q_{\mathcal{F}}$ (as opposed to the literal molecule types). See Fig. \ref{fig9}(i) for an illustration of how a single variable gadget fits in amidst the overall construction of $\Q_{\mathcal{F}}$, and Fig. \ref{fig9}(ii) for the entire construction.

This completes the construction description. Since the size of each gadget is constant, our construction is linear in the size of the Formula $\mathcal{F}$ and therefore polynomial-time computable. We now prove the following claim: $\mathcal{F}$ admits a satisfying assignment if and only if $\Q_{\mathcal{F}}$ has a closed RAF $R'$ with $\hat{x} \not\in {\rm cl}_{R'}(F)$.

\textit{Proof.} ($\implies$) Suppose $\Q_{\mathcal{F}}$ has a closed RAF $R'$ and $\hat{x} \not\in {\rm cl}_{R'}(F)$. Then we construct an assignment $\mathcal{A}$ as follows: for each variable $v \in \mathcal{F}$, let $\mathcal{A}(v) = 1$ if and only if $v^R \in R'$ (set $\mathcal{A}(v) = 0$ otherwise). By (\ref{varCond}),  $\mathcal{A}$ is well-defined. By (\ref{irrP4}), we have $ \bigwedge_{i}{\mathcal{A}(l_{i1}) \vee \mathcal{A}(l_{i2}) \vee \mathcal{A}(l_{i3})}$. It follows that $\mathcal{A} \models \mathcal{F}$.

($\Longleftarrow$) Suppose $\mathcal{F}$ admits a satisfying assignment $\mathcal{A}$. We construct a closed RAF $R'$ with $\hat{x} \not\in {\rm cl}_{R'}(F)$ in four simple steps: (i) Let $r_{in}, r_{out} \in R'$. (ii)
For each variable $v \in \mathcal{F}$ with $\mathcal{A}(v) = 1$, let $v^R \in R'$.  (iii)
For each variable $v \in \mathcal{F}$ with $\mathcal{A}(v) = 0$, let $\ovl{v}^R \in R'$. (iv)
For each positive-valued literal $l_{ij} \in \mathcal{F}$ (with respect to $\mathcal{A}$), let $l_{ij}^R \in R'$.

We now prove that $R'$ is a closed RAF that does not produce $\hat{x}$. We first show that $R'$ is F-generated. To do this, we simply provide the linear ordering of reactions $r_{in}, V, L, r_{out}$ for $V$ and $L$ the variable and literal reactions included in $R'$, respectively.
Besides $r_{in}$, each reaction produces a molecule that catalyses it. Since $r_{in}$ is catalysed by $x_{out}$, and $x_{out}$ is produced by $r_{out}$, we have $r_{in}$ catalysed, so $R'$ is auto-catalytic. 
We now prove $R'$ is closed by case analysis. Suppose $r \in R - R'$, then we have the following cases:

{\em Case (i):} $r = l_{ij}^R \not\in R'$ is a literal reaction. Suppose WLOG that $l_{ij} \in \mathcal{F}$ is a positive occurrence of the variable $v \in \mathcal{F}$. Since $l_{ij}^R \not\in R'$, by the construction of $R'$ we must have $l_{ij}$ negative-valued with respect to $\mathcal{A}$, so $\mathcal{A}(v) = 0$. With $\mathcal{A}(v) = 0$ we necessarily have, by construction, $v^R \not\in R'$. As $v^R$ is the only reaction producing $l_{ij}^X$, we have $l_{ij}^X \not\in {\rm cl}_{R'}(F)$, so $l_{ij}^R$ does not follow by closure. The argument is symmetric for negative literal occurrences in $\mathcal{F}$.

{\em Case (ii):}  $r = v^R \notin R'$ (or $r = \ovl{v}^R \notin R'$) is a variable gadget reaction. By the construction of the variable gadget, its (only) catalysing molecule $v_{cat}^X$ (or $\ovl{v}_{cat}^X$) is not in the closure ${\rm cl}_{R'}(F)$. So $v^R$ ($\ovl{v}^R$) does not follow by closure.

{\em Case (iii):}   $r = v_{mid}^R$ (i.e. a variable gadget reaction producing $\hat{x}$). Since $\mathcal{A}$ is well defined, by construction we cannot have both $v^R, \ovl{v}^R \in R'$, so the reactants $v_{mid}^X, \ovl{v}_{mid}^X$ of $v_{mid}^R$ cannot both be in ${\rm cl}_{R'}(F)$. It follows $v_{mid}^R$ does not follow by closure. This enumerates all cases for reactions not included in $R'$, so $R'$ is closed.

Finally, since $v_{mid}^R \not\in R'$ for all variables $v \in \mathcal{F}$, we clearly have $\hat{x} \not\in {\rm cl}_{R'}(F)$. This completes the proof.
\hfill$\Box$

\bigskip

\bigskip

{\bf Proof of Theorem~\ref{elthm}:}
Our proof will follow a reduction from 3SAT. Given a 3SAT formula $\mathcal{F}$ with $m$ clauses and $n$ variables, we construct in polynomial-time an elementary CRS $\Q_\mathcal{F}^I$ and an inhibiting set $I \subseteq X \times R$ that has a uRAF if and only if $\mathcal{F}$ admits a satisfying assignment.

The construction of $\Q_\mathcal{F}^I = (X, R, C, F)$ from $\mathcal{F}$ is as follows.
 First, we will have a single food molecule $F = \{f\}$. Next, for each clause $c_i = l_{(i,1)} \vee l_{(i, 2)} \vee l_{(i,3)} \in \mathcal{F}$, let $l_{(i,1)}^R, l_{(i,2)}^R, l_{(i,3)}^R \in R$ be a set of `literal' reactions in $R$ and $l_{(i,1)}^X, l_{(i,2)}^X, l_{(i,3)}^X \in X$ be a set of `literal' molecule types in $X$.  For each clause $c_i \in \mathcal{F}$, we will have every literal reaction producing the molecule types corresponding to the next clause $c_{((i+1) \textit{ mod } m) }$ (forming a cycle). Formally, each reaction $l_{(i,j)}^R$ for $0 \leq i < n$ and $1 \leq j \leq 3$ will be defined by:
\\ $l_{(i,j)}^R = (\{f\}, \{l_{((i + 1) \textit{ mod } m, 1)}^X, \: l_{((i + 1) \textit{ mod } m,2)}^X, \: l_{((i + 1) \textit{ mod } m, 3)}^X\})$ with $(l_{(i,j)}^X, l_{(i,j)}^R) \in C$. 

\begin{figure}[ht]
\begin{center}
\includegraphics[width=0.7\linewidth]{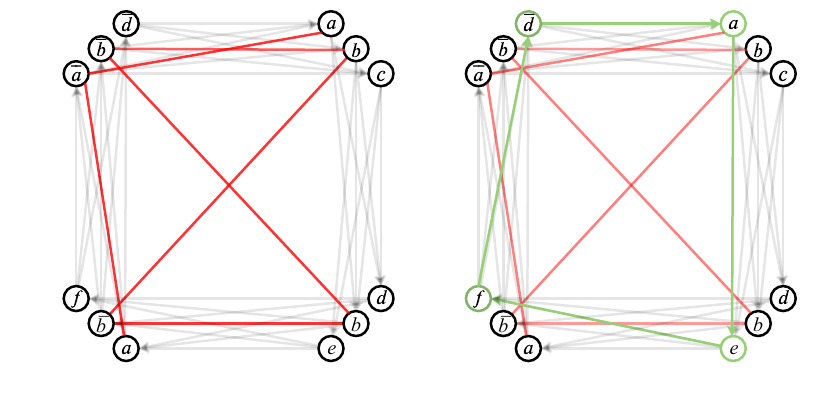} 
  \caption{Given the 3SAT formula $\mathcal{F} = (a \vee b \vee c) \wedge (d \vee b \vee e) \wedge (a \vee \ovl{b} \vee f) \wedge (\ovl{a} \vee \ovl{b} \vee \ovl{d})$, we construct the CRS $\Q_{\mathcal{F}}^I$. The diagram above is simplified version of $\Q_{\mathcal{F}}^I$; each node can be thought of as a literal reaction-molecule pair. The red edges denote the inhibiting relations between pairs of literal reactions / molecule types (enforcing a mutual exclusion) and the grey edges denote the catalytic relations between literal reactions in consecutive clauses. The RAF set (highlighted in green) corresponds to the satisfying assignment with variables $a, e, f$ set true and $\ovl{d}$ false.}
    \label{fig10}
\end{center}
\end{figure}

Finally, for every pair of literal reactions $l_{(i,j)}^R, l_{(s,t)}^R \in R$ which are opposite-parity occurrences of the same underlying variable in $\mathcal{F}$ (see previous proofs), let $(l_{(i,j)}^X, l_{(s,t)}^R) \in I$. See Fig. \ref{fig10} for an abstract illustration.

We now prove that $G_{\mathcal{F}}^I$ has a uRAF if and only if $\mathcal{F}$ admits a satisfying assignment:

($\implies$) Suppose that $G_{\mathcal{F}}^I$ has a uRAF $R' \subseteq R$. We show the existence of a satisfying assignment $\mathcal{A} \models \mathcal{F}$. We define $\mathcal{A}$ to assign variables so that for each literal reaction $l_{(i,j)}^R \in R'$, the literal occurrence $l_{i,j} \in \mathcal{F}$ is positive-valued under $\mathcal{A}$ (i.e. if $l_{i,j} = v$ then $\mathcal{A}(v) = 1$, and if $l_{i,j} = \overline{v}$ then $\mathcal{A}(v) = 0$ for $l_{(i,j)}^R \in R'$). We will let $\mathcal{A}$ assign arbitrarily for literals $l_{(i,j)}$ with $l_{(i,j)}^R \not\in R'$. By construction, since $R'$ is a uRAF,  there cannot simultaneously exist two $l_{(i,j)}^R, l_{(s,t)}^R \in R'$ for $l_{(i,j)}, l_{(s,t)}$ opposite-parity occurrences of the same underlying variable in $\mathcal{F}$ (else $l_{(i,j)}^X \in {\rm cl}_{R'}(F)$ with $(l_{(i,j)}^X,  l_{(s,t)}^R) \in I$ and $R'$ would not be uninhibited). Consequently,  $\mathcal{A}$ is well-defined. Next, as $R' \subseteq R$ is a RAF in $G_{\mathcal{F}}^I$, by construction, every reaction $l_{i,j}^R \in R'$ must be catalysed by $l_{i,j}^X \in {\rm cl}_{R'}(F)$. Since each $l_{i,j}^X$ molecule is produced only by reactions $l_{((i-1) \textit{ mod } m, j')}^R$ for $1 \leq j \leq 3$, it follows that for all $l_{(i,j)}^R \in R'$:
\begin{equation}
l_{(i,j)}^R \in R' \implies l_{((i-1) \textit{ mod } m, 1)}^R \in R' \vee l_{((i-1) \textit{ mod } m, 2)}^R \in R' \vee l_{((i-1) \textit{ mod } m, 3)}^R \in R'
\end{equation}It follows $R'$ must form a cycle with at least one literal reaction $l_{(i,j)}^R$ included per clause in $\mathcal{F}$. By the construction of $\mathcal{A}$ we have every clause positive-valued, so $\mathcal{A} \models \mathcal{F}$.

($\Longleftarrow$)
Suppose that $\mathcal{A} \models \mathcal{F}$ is a satisfying assignment. We construct a uRAF $R'$ as follows: from each clause $c_i \in \mathcal{F}$, arbitrarily select exactly one positive-valued literal $l_{(i,j)} \in c_i$ under $\mathcal{A}$ and include the corresponding reaction $l_{(i,j)}^R$ in the reaction set $R'$. Since exactly one literal reaction is chosen per clause,  each literal reaction $l_{(i,j)}^R \in R'$ is catalysed by $l_{(i,j)}^X\in {\rm cl}_{R'}(F)$ produced by a reaction $l_{((i-1) \text{ mod } m), j')}^R \in R'$ for some $1 \leq j' \leq 3$. Further, since we have no contradictory inclusions of literal reactions (as $\mathcal{A}$ is well-defined), it follows by construction that no reactions are inhibited. Therefore $R'$ is a uRAF.

Finally, as we include only a linear number of reactions, molecule types and catalysis edges in $\mathcal{F}$, we can construct $G_{\mathcal{F}}^I$ in polynomial time.
This completes the proof.
\hfill$\Box$

\end{document}